\documentclass[12pt]{iopart}
\pdfoutput=1
\usepackage[latin9]{inputenc}
\usepackage{color}
\usepackage{verbatim}
\usepackage{iopams}

\usepackage{graphicx}
\usepackage{wasysym}
\usepackage{esint}
\usepackage{array}
\newcolumntype{L}{>{$}l<{$}}
\usepackage[unicode=true,pdfusetitle,
 bookmarks=true,bookmarksnumbered=false,bookmarksopen=false,
 breaklinks=false,pdfborder={0 0 1},backref=false,colorlinks=false]
 {hyperref}
\hypersetup{
 colorlinks,linkcolor=blue,citecolor=blue,urlcolor=blue}

\makeatletter
\pdfoutput=1

\usepackage{mathrsfs}

\usepackage[normalem]{ulem}

\newcommand{\be}{\begin{equation}}
\newcommand{\ee}{\end{equation}}
\newcommand{\bea}{\begin{eqnarray}}
\newcommand{\eea}{\end{eqnarray}}
\makeatother

\begin{document}
\submitto{jopt}
\title{Scattering of graphene plasmons at abrupt interfaces: an analytic and numeric study}

\author{A. J. Chaves$^{1}$, B. Amorim$^{2}$,  Yu. V. Bludov$^{1}$, 
P.~A.~D.~Gonçalves$^{3,4}$, N. M. R. Peres$^{1}$}

\address{$^{1}$Center and Departament of Physics,  and QuantaLab, University of Minho, Campus of Gualtar, 4710-057 Braga, Portugal}

\address{$^{2}$CeFEMA, Instituto Superior Técnico, Universidade de Lisboa,
	Av. Rovisco Pais, 1049-001 Lisboa, Portugal}

\address{$^{3}$Department of Photonics Engineering and Center for Nanostructured 
Graphene, Technical University of Denmark, 
DK-2800 Kgs. Lyngby, Denmark}

\address{$^{4}$Centre for Nano Optics, University of Southern Denmark, 
Campusvej 55, DK-5230 Odense M, Denmark}

\begin{abstract}
We discuss the scattering of graphene
surface plasmon-polaritons (SPPs) at an interface between two semi-infinite
graphene sheets with different doping levels and/or different underlying
dielectric substrates. We take into account retardation effects and
the emission of free radiation in the scattering process. We derive
approximate analytic expressions for the reflection and the transmission
coefficients of the SPPs as well as the same quantities for the emitted
free radiation. We show that the scattering problem can be recast
as a Fredholm equation of the second kind. Such equation can then
be solved by a series expansion, with the first term
of the series correspond to our approximated analytical solution for
the reflection and transmission amplitudes. We have found that almost
no free radiation is emitted in the scattering process and that under
typical experimental conditions the back-scattered SPP transports
very little energy. This work provides a theoretical description of 
graphene plasmon scattering at an interface between distinct Fermi levels 
which could be relevant for the realization of plasmonic circuitry 
elements such as plasmonic lenses or reflectors, and for controlling 
plasmon propagation by modulating the potential landscape of graphene.
\end{abstract}
\maketitle

\section[Introduction]{Introduction}

Controlling the propagation of graphene surface plasmon-polaritons
(SPPs) \cite{booknuno,primer,AsgerRev} is an important technological
problem for applications in SPP circuitry \cite{Pile,Ballester}.
It is well known from elementary wave mechanics that any wave will
be both reflected and transmitted at an interface where the properties of the propagating
medium change. The situation is no different with graphene SPP in
the presence of a spatial change of graphene's conductivity and/or
dielectric properties of the surrounding media.

\begin{figure}[b]
\centering{}\includegraphics[width=8.5cm]{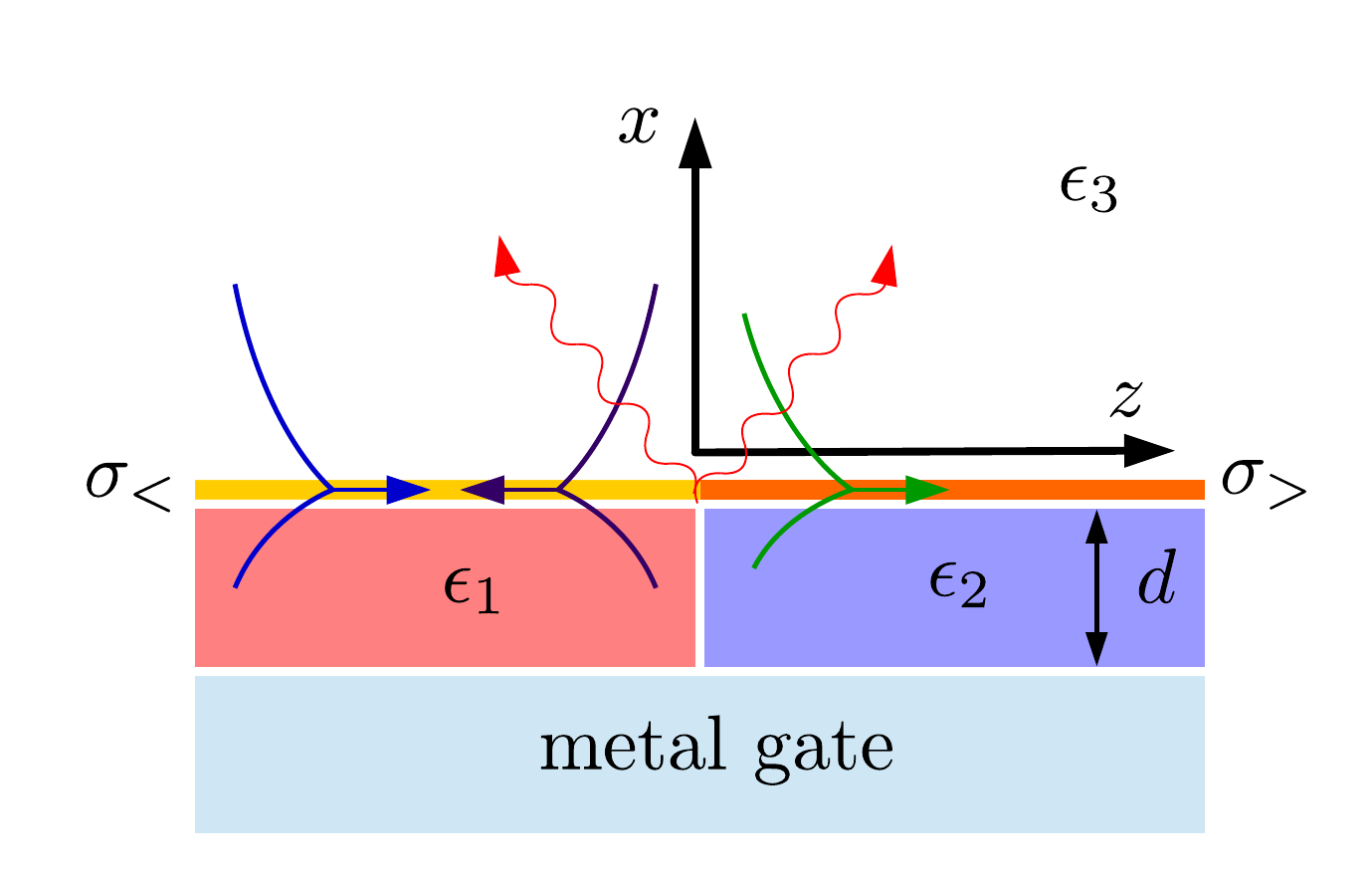} 
\caption[Geometry of the problem]{\label{fig_configuration_1} 
Illustration of the geometry considered for the SPP scattering problem. 
The yellow and red lines stands for graphene at two different electronic 
densities. For simplicity we assume that the electronic density changes 
abruptly at $z=0$, in a step-like manner. We allow for different
dielectric substrates in the regions $z\lessgtr0$. The presence of a metallic
gate allows the tunning of the doping level of the graphene layer. 
A typical SPP scattering event is represented: 
a SPP impinging from the left at the interface can both be reflected and  
transmitted as a SPP, or scattered into free radiation. 
}
\end{figure}

The possibility of generating interfaces for the reflection of graphene
SPP by changing graphene's conductivity is particularly attractive
for the construction of tunable graphene SPP-based circuitry elements,
such as reflectors and beam-splitters, due to the possibility of controlling
graphene's doping level. In a graphene field effect transistor, the doping
of the system is controlled by the gate voltage and by the dielectric
between graphene and the gate electrode \cite{rgraphene1,rgraphene2}.
Therefore, a possible way to create a conductivity interface is to use
a graphene field effect transistor with two different dielectric substrates
below the graphene layer, as depicted in figure \ref{fig_configuration_1}.
Due to the different local capacitances, different electronic densities
will be induced in the two graphene regions, which in turn implies
a different optical conductivity for the two regions. Other possibility
is to consider a single dielectric as the graphene substrate, but
using a split gate geometry, such that the applied gate voltage can
be independently controlled in two different regions \cite{doping}. 
A spatial modulation of graphene's doping level could also be achieved
via non-uniform chemical doping. In general, a graphene SPP incident in a 
conductivity/dielectric interface  will be partially transmitted and partially
reflected. Once the problem of plasmon scattering at a single interface
is solved, it poses no difficulty to create a SPP filter by combining
three different dielectrics in sequence, thereby generalizing the
scheme of the device depicted in figure \ref{fig_configuration_1}.
It should be noted the scattering of a SPP at an interface involves not 
only the transmission and reflection of the field as SPP, but also 
the emission of free radiation\cite{Maradudin0,Maradudin1}.
Ideally, one would want this emission of radiation to be as small as possible
in order to keep the energy within the SPP wave. As we shall see ahead, 
under typical experimental conditions, we predict that the losses in the
scattering event via emission of free propagating radiation are minute.

In this work we study the scattering of
a graphene SPP at normal incidence by a conductivity and/or dielectric
interface. The scattering problem is treated by expanding the electromagnetic
field in terms of a set of local eigenmodes and then using wave matching
at the conductivity/dielectric interface. This method takes into account
both retardation effects and emission of free radiation. Analytic, approximate
expressions are obtained for the graphene SPP reflection and transmission
coefficients. The approximate solution is compared to a numerical solution
of the wavemathcing problem. It is worthwhile pointing out that the problem of reflection
of graphene SPPs at a conductivity step was previously studied
in Ref.~\cite{Rejaei} employing a fully numerical method, but in
the electrostatic limit, which does not take into account radiation losses.
The problem of reflection at a conductivity interface for non-normal incidence
was studied in Ref.~\cite{Saeed}, also in the electrostatic limit.
The scattering of graphene SPPs by a conductivity barrier/well has
been considered in Ref.~\cite{Moreno}, taking into account retardation
effects in a fully numerical approach. In addition, the reflection of SPP at a graphene
edge was studied in Ref.~\cite{Nikitin_Luis}. Research on graphene
plasmonics is a relatively recent topic \cite{booknuno} and research
on graphene plasmonic circuitry is still in its infancy. We note,
however, that imaging of graphene plasmon scattering on lattice defects \cite{Basov,Fei_2017} 
and corrugations \cite{Slipchenko_2017} has already been reported. 
It is also worthwhile noticing that the experimental study of scattering
of SPP in metals has also been reported in Refs.~\cite{Maradudin0,Smolyaninov,Rotenberg,Gordon_2006} 
and the generation of unidirection SPP beams was reported in Ref.~\cite{Oubo_2015}.
On the theoretical side, the problem of scattering of SPP in metals
by one dimensional defects, such as wires or grooves, has been studied
in Refs.~\cite{Pile,Maradudin1,Maradudin2,Arias,Polanco,Aporvari}.
Finally, the scattering of phonon-polaritons at dielectric interfaces has been
studied in Ref.~\cite{paper_surface_phonon}.

This paper is organized as follows: in section~\ref{sec_geometry} we
define the problem and lay down the general approach to tackle it based
on a local eingenmode expansion of the electromagnetic field and wave
matching.  We describe
the electromagnetic mode structure and dispersion relations, considering graphene SPP, waveguide and free radiation
modes. Section~\ref{sec_scattering} is devoted to the problem of graphene
SPP scattering. In section~\ref{subsec_scattering_analytical}, we solve
the scattering problem analytically 
in the approximation of weak coupling of SPPs to radiation modes;
in section~\ref{subsec_scattering_Fredholm} we show that the scattering
problem can be recast as a Fredholm equation of the 
second kind. We show that the approximate results can be 
recovered from the zeroth order solution of the Fredholm equation in section~\ref{subsubsec_recovery_analytical}. 
We compare the analytical results with a numeric solution of the Fredholm equation and discuss the obtained results 
in section~\ref{sec_results}. 
Conclusions are drawn in section~\ref{sec_conclusions}.

\section[Geometry and modes]{Geometry and electromagnetic modes\label{sec_geometry}}

The scattering problem and the geometry we discuss in this work is represented 
in figure \ref{fig_configuration_1}. An identical geometry has been considered 
in the case of scattering of surface phonon-polaritons 
\cite{paper_surface_phonon}.
We assume a plasmon propagating from the left at normal incidence, that is, along the 
$z-$axis. When impinging at the interface between the dielectrics 
$\epsilon_1$ and $\epsilon_2$, part of the plasmon will be reflected, part
will be transmitted, and some of the energy will be radiated to the far field.
We assume a time dependence of the electromagnetic fields of the form
$e^{i\omega t}$.

We obtain the electromagnetic modes of the fields in the geometry depicted in 
figure \ref{fig_configuration_1} by solving  Maxwell's equations 
(see \ref{maxwell_solution}). The resulting
modes are labeled by an index $n$. The properties of these modes are analysed 
in detail in this section. We make a piecewise decomposition of
the fields in terms of the eigenmodes, 
using the superscript $<$($>$) for the $z<0$($z>0$) region:  
\begin{equation}
B_{y}^{\lessgtr}(x,z)=\sum_{n,\lambda} \alpha_{n,\lambda}^{\lessgtr}e^{\lambda 
iq_{n}^{\lessgtr}z}h_{n}^{\lessgtr}(x),\label{eq_expansion_B_2}
\end{equation}
\begin{equation}
E_{x}^{\lessgtr}(x,z)=-\sum_{n,\lambda}\lambda \alpha_{n,\lambda}^{\lessgtr} e^{\lambda 
iq_{n}^{\lessgtr}z}e_{n}^{\lessgtr}(x),\label{eq_expansion_BE_2}
\end{equation}
where $q_n^{\lessgtr}$ is the wavenumber of mode $n$ along the $z$ direction,
$\lambda=\pm 1$ indicates a left/right propagating wave and 
$\alpha_{n,\lambda}^{\lessgtr}$ are mode amplitudes.
We clarify that the sum over $n$ actually denotes a summation over discrete 
modes and an integration over continuum modes.
From \ref{maxwell_solution} the eigenmodes of the  $y$ component of the 
magnetic field read:
\begin{equation}
h_{n}^{\lessgtr}(x)  =
\left\{ \begin{array}{@{\kern2.5pt}lL}
    \hfill  B_{n}^{\lessgtr}e^{p_{3||n}^{\lessgtr}x}+C_{n}^{\lessgtr}e^{-p_{3||n}^{\lessgtr}x}  & if $x>0$,\\
    \hfill  A_{n}^{\lessgtr}\cosh\left[p_{j|n}\left(x+d\right)\right]   & if $0>x>-d$.
\end{array}\right. \label{eq_B_field_graphene_open_2}
\end{equation}
where $A_{n}^{\lessgtr}$, $B_{n}^{\lessgtr}$ and $C_{n}^{\lessgtr}$ are constants to the later defined,
the graphene layer is located at $x=0$ and the metallic gate at $x=-d$, we have written $j=1,2$  
for the  $z<0$, $z>0$ regions, respectively, and
for each region, the wavenumber  along the $x$ direction is given by
\begin{eqnarray}
\left(p_{1|n}\right)^{2}=\left(q_{n}^{<}\right)^{2}-\epsilon_{1}k_{0}{}^{2},\nonumber\\
\left(p_{2|n} \right)^{2}=\left(q_{n}^{>}\right)^{2}-\epsilon_{2}k_{0}^{2},\nonumber\\
\left(p_{3|n}^{\lessgtr}\right)^{2}=\left(q_{n}^{\lessgtr}\right)^{2}-\epsilon_{3}k_{0}^{2}, \label{eq_wavenumbers}
\end{eqnarray}
with $k_{0}=\omega/c$ denoting the wavenumber in vacuum. The relation 
between the wavenumber $q_{n}^{\lessgtr}$ 
and the frequency $\omega$ needs to be calculated for each mode, usually by 
solving a transcendental equation. 
In each region $z\lessgtr0$, the magnetic $h_{n}^{\lessgtr}$ modes can be chosen to satisfiy the 
orthonormality condition (provided there are no losses, i.e. or purely real dielectric functions and a purely imaginary 
graphene conductivity)
\begin{equation}
\langle h_n^{\lessgtr}, e_m^{\lessgtr} \rangle= \int_{-d}^\infty dx \,  h_n^{\lessgtr}(x) e_m^{\lessgtr}(x) 
= \delta_{n,m}, \label{eq_inner_definition}
\end{equation}
where $e_m^{\lessgtr}(x)$ gives the $x$ component of the electric field for mode $m$ (see \ref{maxwell_solution}).

From the boundary conditions on the graphene layer, $x=0$, we obtain the following equations for $A_{n}^{\lessgtr}$,
$B_{n}^{\lessgtr}$ and $C_{n}^{\lessgtr}$
\begin{eqnarray}
B_{n}^{\lessgtr}-C_{n}^{\lessgtr} =\frac{p_{j|n}\epsilon_{3}}{p_{3|n}^{\lessgtr}\epsilon_{j}}A_{n}^{\lessgtr}
\sinh\left(p_{j|n}d\right),\label{eq_graphene_open_boundary_E_2}\\
B_{n}^{\lessgtr}+C_{n}^{\lessgtr}-A_{n}^{\lessgtr}\cosh\left(p_{j|n}d\right)  
=\frac{\sigma_{\lessgtr}}{i\omega\epsilon_{0}}\frac{p_{j|n}}{\epsilon_{j}}A_n^{\lessgtr}\sinh\left(p_{j|n}
d\right).\label{eq_graphene_open_boundary_B_2}
\end{eqnarray}
The solution of the above equations determines the spectrum and the
structure of the electromagnetic modes of the system. The
wavenumbers $p_{j|n}$ and $p_{3|n}^{\gtrless}$ can be real or
purely imaginary. From these possibilities we can classify the modes as: 
graphene SPP (both $p_{j|n}$ and $p_{3|n}^{\gtrless}$ are real), 
waveguide modes ($p_{j|n}$ is imaginary and $p_{3|n}^{\gtrless}$ is real)
and free radiation modes (both $p_{j|n}$ and $p_{3|n}^{\gtrless}$ are imaginary). 
As we do not want to discuss the decay of the modes as they propagate,
but only the scattering event at the interface, we will neglect losses. 
In particular, we neglect the real part of the graphene conductivity,
which we model within a Drude model, by approximating 
$\sigma^{\lessgtr}\simeq i\sigma_{I}^{\lessgtr}$ where
\begin{equation}
\sigma_{I}^{\lessgtr}\simeq-\frac{e^{2}}{\pi\hbar}\frac{E_{F}^{\lessgtr}}{\hbar\omega},\label{eq_Drude_noloss}
\end{equation}
and assume the dielectric constants to be real valued.

\subsection{Graphene SPP}

The graphene SPP is a mode localized in the graphene layer. Further
in the paper we will denote the graphene SPP mode by index $n=0$.
It is characterized by real $p_{j|0}$ and $p_{3|0}^{\lessgtr}$.
The fact that $p_{3|0}^{\lessgtr}$ is real, forces us to set $B_{0}^{\lessgtr}$
to zero in equations ~\ref{eq_graphene_open_boundary_E_2} and \ref{eq_graphene_open_boundary_B_2}, 
in order to avoid the unphysical situation of the field growing
exponentially when $x\rightarrow+\infty$. This leads to the following implicit condition for the 
graphene SPP dispersion
relation
\begin{equation}
\frac{\epsilon_{3}}{p_{3|0}^{\lessgtr}}+\frac{\epsilon_{j}}{p_{j|0}}\coth\left(p_{j|0}d\right)-i\frac{\sigma^{\lessgtr}}{\epsilon_{0}\omega}=0.
\label{eq_dispersion_SPP}
\end{equation}
Clearly, when $d\rightarrow\infty$ we recover the dispersion relation
of plasmons in a graphene layer clad between two semi-infinite dielectrics \cite{booknuno}.
%
We fixe $A_0$ by imposing the normalization condition
\begin{equation}
\left\langle h_{0}^{\lessgtr},e_{0}^{\lessgtr}\right\rangle
=\int_{-d}^{\infty}dx\,h_{0}^{\lessgtr}(x)e_{0}^{\lessgtr}(x)=1,\label{eq_normalization_SPP}
\end{equation}
leading to
\begin{equation}
\left(A_{0}^{\lessgtr}\right)^{2}  =\frac{\omega\epsilon_{j}}{4q_{0}^{\lessgtr}c^{2}}\left(2d+\frac{\sinh\left(2p_{j|0}d\right)}{p_{j|0}}-\frac{\epsilon_{j}}{\epsilon_{3}}\frac{2\epsilon_{0}^{2}\epsilon_{3}^{2}\omega^{2}\cosh^{2}\left(p_{j|0}d\right)}{p_{3|0}^{\lessgtr}\left(p_{3|0}^{\lessgtr}\sigma_{I}^{\gtrless}+\epsilon_{0}\epsilon_{3}\omega\right)^{2}}\right)^{-1}.
\end{equation}

\paragraph{Approximate solution for graphene SPP dispersion relation}

In the electrostatic limit ($c\rightarrow\infty$), we approximate
$p_{j|0}\simeq p_{3|0}^{\gtrless}\simeq q_{0}^{\gtrless}$. With this approximation
equation~\ref{eq_dispersion_SPP} becomes
\begin{equation}
\frac{\epsilon_{3}}{q}+\frac{\epsilon_{j}}{q}\coth\left(qd\right)
-i\frac{\sigma^{\lessgtr}}{\epsilon_{0}\omega}=0.
\end{equation}
Using \ref{eq_Drude_noloss}, we can solve the previous equation for $\omega$ obtaining
\begin{equation}
\hbar\omega=\sqrt{4\alpha E_{F}\hbar c\frac{q}{\epsilon_{3}+\epsilon_{j}\coth\left(qd\right)}},
\label{eq_spp_dispersion}
\end{equation}
where we have introduced the fine-structure constant $\alpha=e^2/(4\pi\epsilon_0c\hbar)$.
In the limit of a thick substrate $qd\gg1$, we approximate $\coth(qd)\simeq 1$, recoverying the 
dispersion relation for a surface plasmon-polariton in graphene supported by an infinite dieletric
\begin{equation}
\hbar\omega\simeq\sqrt{4\alpha E_{F}\hbar c\frac{q}{\epsilon_{3}+\epsilon_{j}}},
\label{eq_spp_thick_substrate}
\end{equation}
with the characteristic $\propto \sqrt{q}$ dependence.
In the opposite limit, $qd\ll1$, we approximate $\coth(qd)\simeq 1/(qd)$ and obtain
\begin{equation}
\hbar\omega\simeq\sqrt{4\alpha E_{F}\hbar c\frac{d}{\epsilon_{j}}}q,
\label{eq_spp_thin_substrate}
\end{equation}
and we obtain a linear dispersion relation for small wavenumbers.

\begin{figure}[h!]
\centering{}
\includegraphics[width=0.6\textwidth]{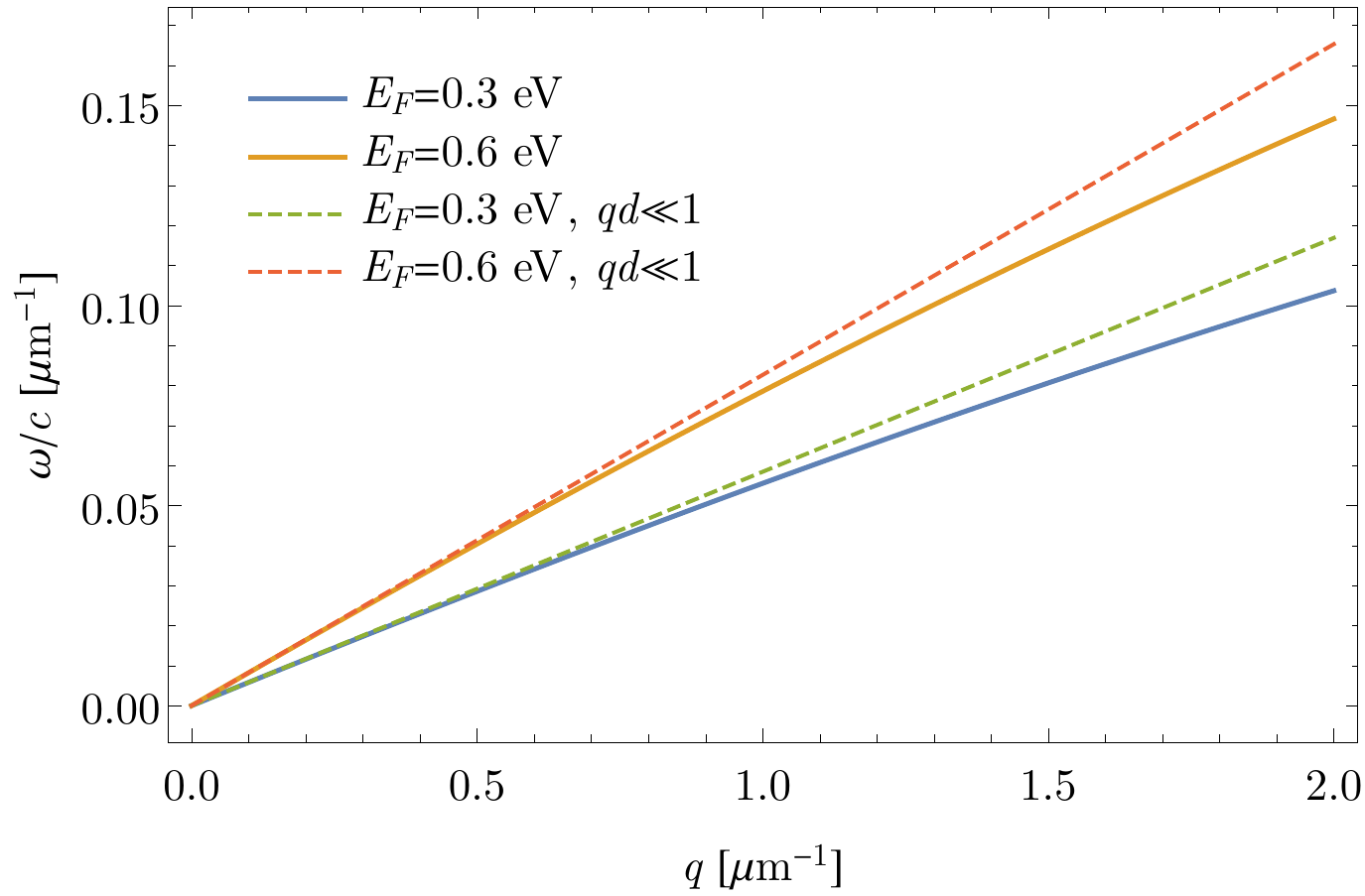}
\caption[Dispersion]{\label{fig_dispersion} 
Dispersion relation of graphene surface plasmon-polariton \ref{eq_spp_dispersion} 
for two different Fermi energies: 
$E_F=0.3$ eV (solid blue line) and $E_F=0.6$ eV (solid yellow line). 
Also represented are the small wavenumber approximations \ref{eq_spp_thin_substrate} for the
plasmon dispersion relation (dashed lines).
The values used in the plot are $d=300$ nm, $\epsilon_j=3.9$ and $\epsilon_3=1$.
}
\end{figure}

In figure \ref{fig_dispersion} we show the dispersion relation of the SPP for
two different Fermi energies. It is clear that for typical substrate tickness and wavenumbers,
the dispersion relation is closer to linear than to the square root dependence.

\subsection{Waveguide modes}

\begin{figure}
\begin{centering}
\includegraphics[width=0.6\textwidth]{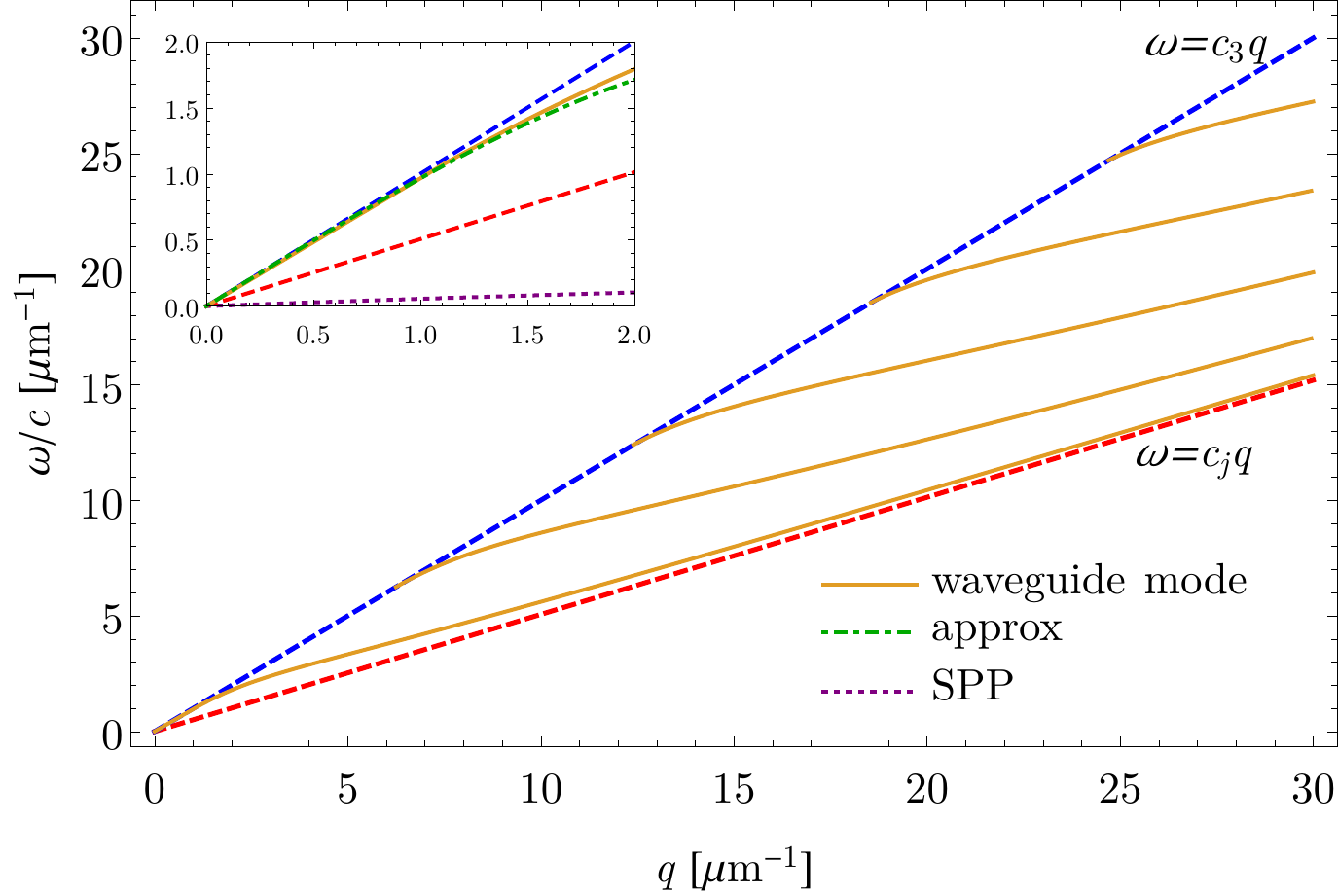}
\par\end{centering}
\caption{\label{fig_Waveguide_dispersion}Dispersion relation, $\omega\left(q_{n}^{<}\right)$, 
(solid yellow lines) for the first five waveguide modes for 
a structure with $d=300$ nm, $\epsilon_j=3.9$, $\epsilon_3=1$, and $E_F=0.3$ eV. The dispersion relation
for the waveguide modes in the absence of graphene is indistinguishable from the dispersion shown on the scale used
The light-lines  $\omega=c_n q$, with $c_n=c/\sqrt{\epsilon_n}$ are shown for 
$\epsilon_n=\epsilon_3$ (blue dashed line) and $\epsilon_n=\epsilon_j$ (red dashed line).
Inset: Zoom in the region with $q$ from $0$ to $2~\mu \rm{m}^{-1}$. The dispersion relation of
the graphene surface plasmon polarion is shown by the dotted purple line, and the 
approximated  dispersion relation for the $n=1$ waveguide mode \ref{eq_lowest_waveguide_approx}
is represented by the dot-dashed green line.
}
\end{figure}

In the case where $\epsilon_{j}>\epsilon_{3}$, the structure supports
modes which are localized in the region $0>x>-d$, dubbed waveguide modes.
Waveguide modes are oscillating in the $0>x>-d$ region, but decay exponentially
for $x\rightarrow\infty$. As in the case for graphene SPP, $p_{3|n}^{\gtrless}$
is real and thus we set $B_{n}^{\lessgtr}=0$. However, due to the
oscillating nature of the field for $0>x>-d$, $p_{j|n}=ik_{j|n}=i\sqrt{\epsilon_{j}k_{0}^{2}-\left(q_{n}^{\lessgtr}\right)^{2}}$
is now purely imaginary. The dispersion relation of the waveguide
modes is still given by equation~\ref{eq_dispersion_SPP}, but with imaginary
$p_{j|n}=ik_{j|n}$. Namely, we obtain the condition
\begin{equation}
\frac{\epsilon_{3}}{p_{3|n}^{\lessgtr}}-\frac{\epsilon_{j}}{k_{j|n}}\cot\left(k_{j|n}d\right)-i\frac{\sigma^{\lessgtr}}{\epsilon_{0}\omega}=0.\label{eq_waveguide_mode_condition}
\end{equation}
The solutions for this equation are organized as a
series of bands with discrete spectrum, $\omega(q_{n}^{\gtrless})$,
restricted to the region 
$cq_{n}^{\gtrless}/\sqrt{\epsilon_{j}}<\omega(q_{n}^{\gtrless})<cq_{n}^{\gtrless}/\sqrt{\epsilon_{3}}$,
as it is shown in figure \ref{fig_Waveguide_dispersion}
for a typical setup. As it can be seen from the figure, the lowest, $n=1$,
waveguide mode bifurcates from the origin and exists for all positive $\omega$ and
$q_{1}^{<}$, while the remaining modes waveguide modes, $n>1$, bifurcate
from the points with frequencies $\omega_{j|n}=c(\pi/d)(n-1)/\sqrt{\epsilon_j/\epsilon_3-1}$,
lying on the light-line in vacuum $\omega=cq/\sqrt{\epsilon_3}$ and existing in the spectral range
above those frequencies, $\omega\ge\omega_{j|n}$. 
The presence of the graphene has a negligible influence on the
spectrum of the waveguide modes for the parameters considered. 


\paragraph{Approximate dispersion relation for the lowest waveguide mode}

In the limit of small frequency and momentum, and neglecting the effect
of the graphene layer, it is possible to obtain an approximate expression
for the lowest, $n=1$, waveguide mode dispersion. Neglecting the graphene
conductivity term in equation~\ref{eq_waveguide_mode_condition} and
approximating $\tan\left(k_{j|1}d\right)\simeq k_{j|1}d$, we obtain
the following condition 
\begin{equation}
1=\frac{\epsilon_{3}k_{j|1}^{2}d}{\epsilon_{j}p_{3|1}^{\lessgtr}}.
\end{equation}
Recalling the definitions of $k_{j|1}=\sqrt{\epsilon_{j}k_{0}^{2}-q_{1}^{\lessgtr2}}$
and $p_{3|1}^{\lessgtr}=\sqrt{q_{1}^{\lessgtr2}-\epsilon_{3}k_{0}^{2}}$,
the previous equation can be solved to lowest order in $q_1^{\lessgtr}$, leading
to the approximate dispersion relation for the $n=1$ waveguide mode 
\begin{equation}
\omega(q_{1}^{\lessgtr})\simeq\frac{cq_{1}^{\lessgtr}}{\sqrt{\epsilon_{3}}}\sqrt{1-\left(q_{1}^{\lessgtr}d\right)^{2}\left(\frac{\epsilon_{j}-\epsilon_{3}}{\epsilon_{j}}\right)^{2}}.
\label{eq_lowest_waveguide_approx}
\end{equation}
This approximate expression for the dispersion relation of the lowest
waveguide mode is shown in figure \ref{fig_Waveguide_dispersion}.
It is clearly seen that for the parameters of figure \ref{fig_Waveguide_dispersion}
this approximation is valid for  $q_{1}^{\lessgtr}\apprle1.5\,\mu\mathrm{m}^{-1}$
and fails for larger wavenumbers.

\subsection{Radiative modes}

Besides localized modes (SPP and waveguide), there is a
continuum of radiative modes. Radiative modes are characterized
by $\epsilon_{3}k_{0}^{2},\epsilon_{j}k_{0}^{2}>\left(q_{n}^{\lessgtr}\right)^{2}$.
We chose to label these modes by their frequency, $\omega$, and
momentum along the $x$ direction in the region $x>0$, $k$, which
we can choose to be positive, such that $p_{3|k}^{\gtrless}=ik$. In this situation, we obtain 
\begin{equation}
q_{k}^{\lessgtr}  =  \sqrt{\epsilon_{3}k_{0}^{2}-k^{2}},\label{eq:radiative-qk}\
\end{equation}
and substituing in \ref{eq_wavenumbers}:
\begin{equation}
p_{j\mathrm{|}k}  =  \sqrt{\left(\epsilon_{3}-\epsilon_{j}\right)k_{0}^{2}-k^{2}},\label{eq:radiative-pjk}
\end{equation}
where we have substituted the index $n$ by $k$.
Equation~\ref{eq:radiative-qk} corresponds to the dispersion relation
of the radiative modes. The aforementioned positiveness of $k$ results
in the fact that the dispersion relation of these modes lies above the light-line for
a dielectric with $\epsilon_{3}$ (see figure \ref{fig_Waveguide_dispersion}).
Notice that for $k^2>k_c^2=\epsilon_{3}k_{0}^{2}$ the radiative modes are actually 
evanescent waves along the $z$ direction with imaginary $q_k^{\lessgtr}$. Therefore, it is with some abuse of language that 
we refer to them as radiation modes. On the other hand, for $k^2<k_c^2$, $q_k^{\lessgtr}$ is real and we wave a true radiation mode
corresponding to a propagating wave in both the $x$ and $z$ directions.
Both kinds of modes are necessary when
making the mode matching at the interface $z=0$. We also have that
$p_{j\mathrm{|}k}$ is real for $k^{2}<\left(\epsilon_{3}-\epsilon_{j}\right)k_{0}^{2}$
(see equation~\ref{eq:radiative-pjk}), thus describing  evanescent
waves along $x$ direction, in the substrate with dielectric constant
$\epsilon_{j}$; and is imaginary in the opposite situation $k^{2}>\left(\epsilon_{3}-\epsilon_{j}\right)k_{0}^{2}$,
which corresponds to the propagating wave along the $x-$direction in the
substrate (when $\epsilon_{j}>\epsilon_{3}$ waves for any $k$
are of that type). For radiation modes all the coefficients
$A_{k}^{\lessgtr}$, $B_{k}^{\lessgtr}$ and $C_{k}^{\lessgtr}$ in equation~\ref{eq_B_field_graphene_open_2}
are non-zero. Imposing the boundary conditions at the $x=0$ interface (see \ref{maxwell_solution}) we can 
write $B_{k}^{\lessgtr}$ and $C_{k}^{\lessgtr}$,
as
\begin{eqnarray}
B_{k}^{\lessgtr}  =\frac{A_{k}^{\lessgtr}}{2}\left(\mathcal{F}_{k}^{\lessgtr}-i\mathcal{G}_{k}^{\lessgtr}\right), \label{eq_radiative_mode_B_coefficient}\\
C_{k}^{\lessgtr}  =\frac{A_{k}^{\lessgtr}}{2}\left(\mathcal{F}_{k}^{\lessgtr}+i\mathcal{G}_{k}^{\lessgtr}\right), \label{eq_radiative_mode_C_coefficient}
\end{eqnarray}
where we have defined
\begin{eqnarray}
\mathcal{F}_{k}^{\lessgtr} =\cosh\left(p_{j|k}d\right)+\frac{\sigma_{I}^{\lessgtr}}{\omega\epsilon_{0}}\frac{p_{j|k}}{\epsilon_{j}}\sinh\left(p_{j|k}
d\right),\\
\mathcal{G}_{k}^{\lessgtr} =\frac{p_{j|k}\epsilon_{3}}{k\epsilon_{j}}\sinh\left(p_{j|k}d\right).
\end{eqnarray}
The electric and magnetic field modes, can thus be written as
\begin{equation}
h_{k}^{\lessgtr}(x)=A_{k}^{\lessgtr}\left\{ \begin{array}{@{\kern2.5pt}lL} 
\hfill \mathcal{F}_{k}^{\lessgtr}\cos\left(kx\right)+\mathcal{G}_{k}^{\lessgtr}\sin\left(kx\right) & if $x>0$\\
\hfill \cosh\left[p_{j|k}\left(x+d\right)\right] ,& if $0>x>-d$
\end{array}\right.,
\end{equation}
and the corresponding $x$ component of the eletric field reads
\begin{equation}
e_{k}^{\lessgtr}(x)=A_{k}^{\lessgtr}\frac{q_{k}^{\lessgtr}c^{2}}{\omega\epsilon_{3}}\left\{ \begin{array}{@{\kern2.5pt}lL} 
\hfill \mathcal{F}_{k}^{\lessgtr}\cos\left(kx\right)+\mathcal{G}_{k}^{\lessgtr}\sin\left(kx\right) & ,$x>0$\\
\hfill \frac{\epsilon_{3}}{\epsilon_{j}}\cosh\left[p_{j|k}\left(x+d\right)\right] &  ,$0>x>-d$
\end{array}\right..
\end{equation}

The modes can be normalized through the condition:
\begin{equation}
\int_{-d}^{\infty}dx \,h_{k}^{\lessgtr}(x)e_{k^{\prime}}^{\lessgtr}(x)=\delta(k-k^{\prime}),\label{eq_normalization_delta}
\end{equation}
which fixes $A_{k}^{\lessgtr}$ to have the value
\begin{equation}
\left(A_{k}^{\lessgtr}\right)^{2}=\frac{\omega\epsilon_{3}}{q_{k}^{\lessgtr}c^{2}}\frac{2}{\pi}\frac{1}{\left|{\cal F}_{k}^{\lessgtr}\right|^{2}+\left|{\cal G}_{k}^{\lessgtr}\right|^{2}}.
\label{eq_normalization_radiative}
\end{equation}
Notice that $A_k^{\lessgtr}$ will be imaginary when $q_k^{\lessgtr}$ is imaginary.

\section{SPP scattering\label{sec_scattering}}

We now consider the problem of scattering of a graphene SPP which
is illustrated in figure \ref{fig_configuration_1}. A plasmon coming
from the left and impinging at the dielectric/conductivity interface at
$z=0$ is scattered into both a back-scattered (reflected) and 
forward-scattered (transmitted) plasmon, and also into free propagating 
radiation. For simplicity, we will
consider a situation where no waveguide modes are supported ($\epsilon_{j}<\epsilon_{3}$).
In order to determine the total field in the regions $z\lessgtr0$,
we must consider both the discrete plasmon mode and the radiative modes.
Therefore the expansion of the electric and magnetic fields in terms
of local eigenmodes, equations \ref{eq_expansion_B_2} and \ref{eq_expansion_BE_2},
reads for $z<0$ (note the phase of $\pi$ introduced in the reflection coefficients of the electric field)
\begin{eqnarray}
E_{x}^{<}(x,z) & =e_{0}^{<}(x)e^{-iq_{0}^{<}z}-e_{0}^{<}(x)r_{0}e^{iq_{0}^{<}z}  -\int_{0}^{\infty}dk\,r_{k}e_{k}^{<}(x)e^{iq_{k}^{<}z},\label{eq_E_less}\\
B_{y}^{<}(x,z) & =h_{0}^{<}(x)e^{-iq_{0}^{<}z}+h_{0}^{<}(x)r_{0}e^{iq_{0}^{<}z} 
 +\int_{0}^{\infty}dk\,r_{k}h_{k}^{<}(x)e^{iq_{k}^{<}z},\label{eq_B_less}
\end{eqnarray}
while for $z>0$ we write
\begin{eqnarray}
E_{x}^{>}(x,z) & =e_{0}^{>}(x)\tau_{0}e^{-iq_{0}^{>}z}+\int_{0}^{\infty}dk\tau_{k}e_{k}^{>}(x)e^{-iq_{k}^{>}z},\label{eq_Egreater}\\
B_{y}^{>}(x,z) & =h_{0}^{>}(x)\tau_{0}e^{-iq_{0}^{>}z}+\int_{0}^{\infty}dk\tau_{k}h_{k}^{>}(x)e^{-iq_{k}^{>}z}.\label{eq_Bgreater}
\end{eqnarray}
In these expressions, $r_{0}/\tau_{0}$ and $r_{k}/\tau_{k}$ are,
respectively, the reflection/transmission amplitudes for the SPP and
radiative modes with wavenumber $k$ along the $x-$direction, for $x>0$. The relation
between the frequency $\omega$ and the in-plane graphene SPP momentum,
$q_{0}^{\lessgtr}$, is determined by equation~\ref{eq_spp_dispersion}.

Performing mode matching by enforcing the continuity of $E_{x}(x,z)$ and $B_{y}(x,z)$ at $z=0$, we obtain the set of equations
\begin{eqnarray}
e_{0}^{<}(x)(1-r_{0})  -\int_{0}^{\infty}dk\,r_{k}e_{k}^{<}(x)=\tau_{0}e_{0}^{>}(x) 
  +\int_{0}^{\infty}dk\,\tau_{k}e_{k}^{>}(x)\,,\label{eq_matching_1}\\
h_{0}^{<}(x)(1+r_{0})  +\int_{0}^{\infty}dk\,r_{k}h_{k}^{<}(x)=\tau_{0}h_{0}^{>}(x) 
  +\int_{0}^{\infty}dk\,\tau_{k}h_{k}^{<}(x).\label{eq_matching_2}
\end{eqnarray}
Note that in order to satisfy the matching conditions at $z=0$, we need both propagating
and evanescent radiative modes along the $z$ direction.
To determine the reflection and
transmission amplitudes, we take the inner product (as defined in \ref{eq_inner_definition}) of \ref{eq_matching_1} with $h_{0}^{>}(x)$
and $h_{k}^{>}(x)$, and the inner procuct of \ref{eq_matching_2} with $e_{0}^{>}(x)$
and $e_{k}^{>}(x)$. Using the orthonormality
of the modes, we obtain the following system
of equations
\begin{eqnarray}
\tau_{0} & =\left(1-r_{0}\right)\left\langle h_{0}^{>},e_{0}^{<}\right\rangle -\int_{0}^{\infty}dk\,r_{k}\left\langle h_{0}^{>},e_{k}^{<}\right\rangle ,\label{eq_tau_a}\\
\tau_{0} & =\left(1+r_{0}\right)\left\langle e_{0}^{>},h_{0}^{<}\right\rangle +\int_{0}^{\infty}dk\,r_{k}\left\langle e_{0}^{>},h_{k}^{<}\right\rangle ,\label{eq_tau_b}
\end{eqnarray}
and 
\begin{eqnarray}
\tau_{k}= & \left(1-r_{0}\right)\left\langle h_{k}^{>},e_{0}^{<}\right\rangle -\int_{0}^{\infty}dk^{\prime}r_{k^{\prime}}\left\langle h_{k}^{>},e_{k^{\prime}}^{<}\right\rangle ,\label{eq_tauR_a}\\
\tau_{k}= & \left(1+r_{0}\right)\left\langle e_{k}^{>},h_{0}^{<}\right\rangle +\int_{0}^{\infty}dk^{\prime}r_{k^{\prime}}\left\langle e_{k}^{>},h_{k^{\prime}}^{<}\right\rangle .\label{eq_tauR_b}
\end{eqnarray}
The solution of this system of coupled integral equations
yields the reflection and transmission amplitudes. 
In the following, we will provide both an approximate analytic solution 
and a full numerical solution for this system of equations.

\subsection{Approximate analytical solution\label{subsec_scattering_analytical}}

In order to proceed analytically, we will introduce some approximations.
We assume that the following relations hold \cite{paper_surface_phonon}
\begin{eqnarray}
\left\langle h_{0}^{>},e_{k}^{<}\right\rangle   \simeq\left\langle e_{0}^{>},h_{k}^{<}\right\rangle \simeq\left\langle h_{k}^{>},e_{0}^{<}\right\rangle \simeq\left\langle e_{k}^{>},h_{0}^{<}\right\rangle \simeq0,\label{eq_approx_a}\\
\langle h_{k}^{>},e_{k^{\prime}}^{<}\rangle  \simeq\delta\left(k-k^{\prime}\right).\label{eq_approx_b}
\end{eqnarray}
Mathematically these relations mean that the modes of the different
regions are almost orthogonal. Physically, we can
understand this as a statement that the SPP modes are weakly coupled
to the radiation modes. The previous relations
are approximately true as long as $\epsilon_{1}\simeq\epsilon_{2}$
and $E_{F}^{<}\simeq E_{F}^{>}$. 
This regime implies small
reflection amplitudes, as can be seen in figure \ref{fig_transmittance_reflectance}. However, as we
will see below, the approximation performs well even beyond this
regime. With the aforementioned approximations, equations \ref{eq_tau_a} and \ref{eq_tau_b}
become
\begin{eqnarray}
\tau^{\rm{approx}}_{0} & =\left(1-r^{\rm{approx}}_{0}\right)\left\langle h_{0}^{>},e_{0}^{<}\right\rangle ,\\
\tau^{\rm{approx}}_{0} & =\left(1+r^{\rm{approx}}_{0}\right)\left\langle e_{0}^{>},h_{0}^{<}\right\rangle .
\end{eqnarray}
We have thus obtained a closed set of two equations for the SPP reflection
and transmission coefficients. Solving these, we obtain
\begin{eqnarray}
r^{\rm{approx}}_{0} & =\frac{\left\langle h_{0}^{>},e_{0}^{<}\right\rangle -\left\langle e_{0}^{>},h_{0}^{<}\right\rangle }{\left\langle h_{0}^{>},e_{0}^{<}\right\rangle +\left\langle e_{0}^{>},h_{0}^{<}\right\rangle },\label{eq_r_spp_1}\\
\tau^{\rm{approx}}_{0} & =2\frac{\left\langle h_{0}^{>},e_{0}^{<}\right\rangle \left\langle e_{0}^{>},h_{0}^{<}\right\rangle }{\left\langle h_{0}^{>},e_{0}^{<}\right\rangle +\left\langle e_{0}^{>},h_{0}^{<}\right\rangle }.\label{eq_t_spp_1}
\end{eqnarray}
The transmission and reflection coefficients for the radiative modes can
be obtained from equations \ref{eq_tauR_a} and \ref{eq_tauR_b} if
we use the approximation \ref{eq_approx_b}, while keeping $\left\langle h_{k}^{>},e_{0}^{<}\right\rangle $
and $\left\langle e_{k}^{>},h_{0}^{<}\right\rangle $ (in order to
obtain a non-zero result). We obtain the following equations
\begin{eqnarray}
\tau^{\rm{approx}}_{k}= & \left(1-r^{\rm{approx}}_{0}\right)\left\langle h_{k}^{>},e_{0}^{<}\right\rangle -r^{\rm{approx}}_{k},\\
\tau^{\rm{approx}}_{k}= & \left(1+r^{\rm{approx}}_{0}\right)\left\langle e_{k}^{>},h_{0}^{<}\right\rangle +r^{\rm{approx}}_{k}.
\end{eqnarray}
Using the previously obtained value for $r_{0}$, we can solve for
$r_{k}$ and $\tau_{k}$, yielding
\begin{eqnarray}
r^{\rm{approx}}_{k} & =\frac{\left\langle e_{0}^{>},h_{0}^{<}\right\rangle \left\langle h_{k}^{>},e_{0}^{<}\right\rangle -\left\langle h_{0}^{>},e_{0}^{<}\right\rangle \left\langle e_{k}^{>},h_{0}^{<}\right\rangle }{\left\langle h_{0}^{>},e_{0}^{<}\right\rangle +\left\langle e_{0}^{>},h_{0}^{<}\right\rangle },\label{eq_reflection_k_analytic}\\
\tau^{\rm{approx}}_{k} & =\frac{\left\langle e_{0}^{>},h_{0}^{<}\right\rangle \left\langle h_{k}^{>},e_{0}^{<}\right\rangle +\left\langle h_{0}^{>},e_{0}^{<}\right\rangle \left\langle e_{k}^{>},h_{0}^{<}\right\rangle }{\left\langle h_{0}^{>},e_{0}^{<}\right\rangle +\left\langle e_{0}^{>},h_{0}^{<}\right\rangle }.\label{eq_transmission_k_analytic}
\end{eqnarray}
The inner products in the above equations can be computed analytically
and explicit expressions are given in ~\ref{app_inner}. 

One comment regarding the validity of the employed approximations
is in order. Notice that instead of contracting equations \ref{eq_matching_1}
and \ref{eq_matching_2} with $h_{0}^{> }(x)$ and $e_{0}^{> }(x)$,
as done to obtain  equations \ref{eq_tau_a} and \ref{eq_tau_b},
we could have contracted them with $h_{0}^{<}(x)$ and $e_{0}^{<}(x)$.
Such a procedure would lead to the following equations
\begin{eqnarray}
1-r_{0} & =\tau_{0}\left\langle h_{0}^{<},e_{0}^{>}\right\rangle +\int_{0}^{\infty}dk\,\tau_{k}\left\langle h_{0}^{<},e_{k}^{>}\right\rangle ,\\
1+r_{0} & =\tau_{0}\left\langle e_{0}^{<},h_{0}^{>}\right\rangle +\int_{0}^{\infty}dk\,\tau_{k}\left\langle e_{0}^{<},h_{k}^{<}\right\rangle .
\end{eqnarray}
Using the approximations \ref{eq_approx_a} and \ref{eq_approx_b}, 
we obtain
\begin{eqnarray}
1-r^{\rm{approx}^\prime}_{0}  =\tau^{\rm{approx}^\prime}_{0}\left\langle h_{0}^{<},e_{0}^{>}\right\rangle ,\\
1+r^{\rm{approx}^\prime}_{0}  =\tau^{\rm{approx}^\prime}_{0}\left\langle e_{0}^{<},h_{0}^{>}\right\rangle .
\end{eqnarray}
Solving these equations, gives us the alternative expressions for the reflection
and transmission coefficients
\begin{eqnarray}
r^{\rm{approx}^\prime}_{0} & =\frac{\left\langle e_{0}^{<},h_{0}^{>}\right\rangle -\left\langle h_{0}^{<},e_{0}^{>}\right\rangle }{\left\langle e_{0}^{<},h_{0}^{>}\right\rangle +\left\langle h_{0}^{<},e_{0}^{>}\right\rangle },\label{eq_r_spp_2}\\
\tau^{\rm{approx}^\prime}_{0} & =\frac{2}{\left\langle e_{0}^{<},h_{0}^{>}\right\rangle +\left\langle h_{0}^{<},e_{0}^{>}\right\rangle }.\label{eq_t_spp_2}
\end{eqnarray}
Since $e_{0}^{\lessgtr}$ and $h_{0}^{\lessgtr}$ can be chosen
as real, we conclude that equations \ref{eq_r_spp_1} and \ref{eq_r_spp_2}
for $r_{0}$ coincide. However, we see that equations ~\ref{eq_t_spp_1}
and \ref{eq_t_spp_2} differ by a factor of $\left\langle h_{0}^{>},e_{0}^{<}\right\rangle \left\langle e_{0}^{>},h_{0}^{<}\right\rangle $.
This gives us an internal consistency check for the employed approximations:
they remain valid as long as
\begin{equation}
\left\langle h_{0}^{>},e_{0}^{<}\right\rangle \left\langle e_{0}^{>},h_{0}^{<}\right\rangle \simeq1,
\end{equation}
which implies a strong coupling between the SPP modes from $z<0$ and for $z>0$. 

Note that the value for $r_{0}$ obtained with these approximations is 
purely real. Therefore there is no phase-shift in the back-scattering
amplitude of the plasmon, except for the already included phase-shift
of $\pi$. This is a consequence of the approximation introduced above
and contrasts with the results of 
\cite{Rejaei,Nikitin_Luis}, obtained within the electrostatic limit, 
thus ignoring retardation effects.

It should also be noted that the formalism is capable of describing the reflection
of a graphene plasmon at the edge of a semi-infinite graphene sheet.
We have verified numerically that in this case the transmittance is
numerically very small (due to the approximations is not exactly zero)
and the reflectance is essential equal to unity (results not shown;
numerically we take the Fermi energy at the right of $z=0$ a very
small number, typically $E_{F}^>\sim10^{-3}E_{F}^<$, as the numerical procedure
does not allow a zero Fermi energy).

In Ref.~\cite{Rejaei}, an electrostatic calculation predicts that
the reflection coefficient for graphene in vacuum and subject to a
conductivity step at $z=0$ is given by
\begin{equation}
\left|r_{0}\right|^{2}=\left(\frac{q_{0}^{<}-q_{0}^{>}}{q_{0}^{<}+q_{0}^{>}}\right)^{2}.\label{eq_electro}
\end{equation}
If we use the numbers of figure \ref{fig_t_and_r_coefficients} for the Fermi energies
and plug-in the corresponding wavevectors in formula  \ref{eq_electro} 
we obtain the value $\left|r_{0}\right|^{2}\approx0.049$, whereas
our calculation in the same conditions predicts a value in the range
$\left|r_{0}\right|^{2}\approx0.049-0.016$, as the frequency of the
incoming SPP ranges from zero to $\sim$16 meV. Note that a consequence of the electrostatic approximation is
that the reflection coefficient becomes frequency independent. When taking the electrostatic
limit, we can study two possible cases: (i) thin substrate limit,
$d\rightarrow0$, and (ii) thick substrate limit, $d\rightarrow\infty$. 

In the electrostatic and thin substrate limits ($\omega/c,d\rightarrow0$) the reflectance amplitude \ref{eq_r_spp_1} reads 
\begin{equation}
r_{0}=\frac{\epsilon_{2}q_{0}^{<}-\epsilon_{1}q_{0}^{>}}{\epsilon_{2}q_{0}^{<}+\epsilon_{1}q_{0}^{>}},\label{eq_electro_r}
\end{equation}
in agreement with the result of Ref.~\cite{Rejaei} for
$\epsilon_{1}=\epsilon_{2}$. 
For the transmittance amplitude \ref{eq_t_spp_1}, and in the same limit as before, we obtain 
\begin{equation}
\tau_{0}=\frac{2\sqrt{q_{0}^{>}q_{0}^{<}\epsilon_{1}\epsilon_{2}}}{\epsilon_{2}q_{0}^{<}+\epsilon_{1}q_{0}^{>}}.\label{eq_electro_t}
\end{equation}
Physically, the limit $d\rightarrow0$ means that the plasmon fields
are finite only in the dielectric $\epsilon_{3}$, as the field is
screened by the metallic gate. We also note that equations \ref{eq_electro_r}
and \ref{eq_electro_t} contain the limit of total reflection when
$q_{0}^{<}\rightarrow0$. As anticipated, it is possible to have SPP
reflection even if if $E_F^{<}=E_F^{>}$, 
provided that $\epsilon_{1}$ and $\epsilon_{2}$ differ. 

Conversely, in the electrostatic and thick substrate limits ($\omega/c\rightarrow0$,
$d\rightarrow\infty$), we obtain for $r_{0}$ \ref{eq_r_spp_1} and $\tau_{0}$ \ref{eq_t_spp_1}
\begin{eqnarray}
r_{0} & =\frac{\left(\epsilon_{2}+\epsilon_{3}\right)q_{0}^{<}-\left(\epsilon_{1}+\epsilon_{3}\right)q_{0}^{>}}{\left(\epsilon_{2}+\epsilon_{3}\right)q_{0}^{<}+\left(\epsilon_{1}+\epsilon_{3}\right)q_{0}^{>}},\label{eq_electro_r_b}\\
\tau_{0} & =\frac{4q_{0}^{<}q_{0}^{>}\sqrt{\left(\epsilon_{1}+\epsilon_{3}\right)\left(\epsilon_{2}+\epsilon_{3}\right)}}{\left(q_{0}^{<}+q_{0}^{>}\right)\left[\left(\epsilon_{2}+\epsilon_{3}\right)q_{0}^{<}+\left(\epsilon_{1}+\epsilon_{3}\right)q_{0}^{>}\right]}.\label{eq_electro_t_b}
\end{eqnarray}

\subsection{Formulation as a Fredholm equation}\label{subsec_scattering_Fredholm}

We will now recast the scattering problem in a form ameable to a numerical solution. While doing that, we will see how the 
approximate analytic result corresponds to a lowest order approximation to the solution of the complete problem.

Recalling equations \ref{eq_tau_a}-\ref{eq_tauR_b} and
subtracting equation~\ref{eq_tau_b} from equation~\ref{eq_tau_a}, we obtain
\begin{eqnarray}
r_{0}=\frac{\left\langle h_{0}^{>},e_{0}^{<}\right\rangle -\left\langle e_{0}^{>},h_{0}^{<}\right\rangle }{\left\langle h_{0}^{>},e_{0}^{<}\right\rangle +\left\langle e_{0}^{>},h_{0}^{<}\right\rangle }
-\int_{0}^{\infty}dk\frac{\left\langle h_{0}^{>},e_{k}^{<}\right\rangle +\left\langle e_{0}^{>},h_{k}^{<}\right\rangle }{\left\langle h_{0}^{>},e_{0}^{<}\right\rangle +\left\langle e_{0}^{>},h_{0}^{<}\right\rangle }r_{k}.\label{eq_reflection_integral}
\end{eqnarray}
Furthermore, subtracting equation~\ref{eq_tauR_b} from equation~\ref{eq_tauR_a}, we obtain
\begin{eqnarray}
r_{0}=\frac{\left\langle h_{k}^{>},e_{0}^{<}\right\rangle -\left\langle e_{k}^{>},h_{0}^{<}\right\rangle }{\left\langle h_{k}^{>},e_{0}^{<}\right\rangle +\left\langle e_{k}^{>},h_{0}^{<}\right\rangle }
-\int_{0}^{\infty}dk^{\prime}\frac{\left\langle h_{k}^{>},e_{k^{\prime}}^{<}\right\rangle +\left\langle e_{k}^{>},h_{k^{\prime}}^{<}\right\rangle }{\left\langle h_{k}^{>},e_{0}^{<}\right\rangle +\left\langle e_{k}^{>},h_{0}^{<}\right\rangle }r_{k^{\prime}}\label{eq_reflection_integral_2}
\end{eqnarray}
Combining equations \ref{eq_reflection_integral} and \ref{eq_reflection_integral_2}
we eliminate $r_{0}$ and obtain a closed equation for the reflection
coefficients $r_{k}$
\begin{equation}
z_{1}(k)+\int_{0}^{\infty}dk^{\prime}z_{2}(k,k^{\prime})r_{k^{\prime}}=0,\label{eq_Freholm_first}
\end{equation}
where we have introduced the quantities
\begin{equation}
z_{1}(k)=\frac{\left\langle h_{0}^{>},e_{0}^{<}\right\rangle -\left\langle e_{0}^{>},h_{0}^{<}\right\rangle }{\left\langle h_{0}^{>},e_{0}^{<}\right\rangle +\left\langle e_{0}^{>},h_{0}^{<}\right\rangle }
-\frac{\left\langle h_{k}^{>},e_{0}^{<}\right\rangle -\left\langle e_{k}^{>},h_{0}^{<}\right\rangle }{\left\langle h_{k}^{>},e_{0}^{<}\right\rangle +\left\langle e_{k}^{>},h_{0}^{<}\right\rangle },\label{eq_z1}
\end{equation}
\begin{equation}
z_{2}(k,k^{\prime})=\frac{\left\langle h_{k}^{>},e_{k^{\prime}}^{<}\right\rangle +\left\langle e_{k}^{>},h_{k^{\prime}}^{<}\right\rangle }{\left\langle h_{k}^{>},e_{0}^{<}\right\rangle +\left\langle e_{k}^{>},h_{0}^{<}\right\rangle }
-\frac{\left\langle h_{0}^{>},e_{k^{\prime}}^{<}\right\rangle +\left\langle e_{0}^{>},h_{k^{\prime}}^{<}\right\rangle }{\left\langle h_{0}^{>},e_{0}^{<}\right\rangle +\left\langle e_{0}^{>},h_{0}^{<}\right\rangle }.\label{eq_z2}
\end{equation}
Equation~\ref{eq_Freholm_first} is in the form of a Fredholm integral
equation of the first kind. However, as shown in the \ref{app_inner},
the integration kernel $z_{2}(k,k^{\prime})$ contains a term that
is proportional to a $\delta$-function (see equations \ref{eq_inner_delta_1}
and \ref{eq_inner_delta_2}). Therefore, we can split $z_{2}(k,k^{\prime})$
as
\begin{equation}
z_{2}(k,k^{\prime})=v(k)\delta(k-k^{\prime})+v(k)z_{3}(k,k^{\prime}),\label{eq_z3}
\end{equation}
where $v(k)$ is the diagonal part of $z_{2}(k,k^{\prime})$, with its
explicit form given in equation~\ref{eq_v_function_compact}, and we
have written the remaining part as $v(k)z_{3}(k,k^{\prime})$. Inserting
this equation into equation~\ref{eq_Freholm_first} and using the $\delta$-function
to perform the integration over $k^{\prime}$, we can transform the
problem into a Fredholm integral equation of the second kind, as
\begin{equation}
r_{k}=-\frac{z_{1}(k)}{v(k)}-\int_{0}^{\infty}dk^{\prime}z_{3}(k,k^{\prime})r_{k^{\prime}}.\label{eq_Fredholm_second}
\end{equation}
This equation can be solved numerically, by discretizing the integral over $k^\prime$ using a Gaussian quadrature method,
and evaluating the equation for values of $k$ on that same discretized grid, reduzing the integral equation
to a problem of linear algebra as described in greater detail in \ref{numerical_solution}. 

Having obtained the reflection coefficient $r_k$, the reflection coefficient for hte 
the SPP mode, $r_{0}$, can be computed from equation~\ref{eq_reflection_integral}.
With the knowledge of all the reflection coefficients, the transmission
coefficient $\tau_{0}$ can be calculated from equation~\ref{eq_tau_a} as
\begin{eqnarray}
\tau_{0}=2\frac{\left\langle h_{0}^{>},e_{0}^{<}\right\rangle \left\langle e_{0}^{>},h_{0}^{<}\right\rangle }{\left\langle h_{0}^{>},e_{0}^{<}\right\rangle +\left\langle e_{0}^{>},h_{0}^{<}\right\rangle }+\\
+\int_{0}^{\infty}dk\frac{\left\langle h_{0}^{>},e_{0}^{<}\right\rangle \left\langle e_{0}^{>},h_{k}^{<}\right\rangle -\left\langle h_{0}^{>},e_{k}^{<}\right\rangle \left\langle e_{0}^{>},h_{0}^{<}\right\rangle }{\left\langle h_{0}^{>},e_{0}^{<}\right\rangle +\left\langle e_{0}^{>},h_{0}^{<}\right\rangle }r_{k},\label{eq_transmission_integral}
\end{eqnarray}
and the transmission coefficients $\tau_{k}$ can be determined from
equations \ref{eq_reflection_integral} and \ref{eq_tauR_b} as
\begin{eqnarray}
\tau_{k}=\frac{\left\langle h_{0}^{>},e_{0}^{<}\right\rangle \left\langle e_{k}^{>},h_{0}^{<}\right\rangle +\left\langle e_{0}^{>},h_{0}^{<}\right\rangle \left\langle h_{k}^{>},e_{0}^{<}\right\rangle }{\left\langle h_{0}^{>},e_{0}^{<}\right\rangle +\left\langle e_{0}^{>},h_{0}^{<}\right\rangle }+\\
+\int_{0}^{\infty}dk^{\prime}\left[\left\langle e_{k}^{>},h_{k^{\prime}}^{<}\right\rangle -\frac{\left\langle h_{0}^{>},e_{k^{\prime}}^{<}\right\rangle +\left\langle e_{0}^{>},h_{k^{\prime}}^{<}\right\rangle }{\left\langle h_{0}^{>},e_{0}^{<}\right\rangle +\left\langle e_{0}^{>},h_{0}^{<}\right\rangle }\left\langle e_{k}^{>},h_{0}^{<}\right\rangle \right]r_{k^{\prime}}\\
-\frac{\left\langle e_{0}^{>},h_{0}^{<}\right\rangle \left\langle h_{k}^{>},e_{0}^{<}\right\rangle -\left\langle h_{0}^{>},e_{0}^{<}\right\rangle \left\langle e_{k}^{>},h_{0}^{<}\right\rangle }{\left\langle h_{0}^{>},e_{0}^{<}\right\rangle +\left\langle e_{0}^{>},h_{0}^{<}\right\rangle }\label{eq_transmisson_k_integral}
\end{eqnarray}
This provides a general scheme to fully solve the scattering problem.

Notice that
equations \ref{eq_reflection_integral}, \ref{eq_transmission_integral},
and \ref{eq_transmisson_k_integral} can be rewritten as
\begin{equation}
r_{0}=r_{0}^{\mathrm{approx}}-\int_{0}^{\infty}dk\frac{\left\langle h_{0}^{>},e_{k}^{<}\right\rangle +\left\langle e_{0}^{>},h_{k}^{<}\right\rangle }{\left\langle h_{0}^{>},e_{0}^{<}\right\rangle +\left\langle e_{0}^{>},h_{0}^{<}\right\rangle }r_{k},\label{eq_r_0_reconstruction}
\end{equation}
\begin{equation}
\tau_{0}=\tau_{0}^{\mathrm{approx}}
+\int_{0}^{\infty}dk\frac{\left\langle h_{0}^{>},e_{0}^{<}\right\rangle \left\langle e_{0}^{>},h_{k}^{<}\right\rangle -\left\langle h_{0}^{>},e_{k}^{<}\right\rangle \left\langle e_{0}^{>},h_{0}^{<}\right\rangle }{\left\langle h_{0}^{>},e_{0}^{<}\right\rangle +\left\langle e_{0}^{>},h_{0}^{<}\right\rangle }r_{k},\label{eq_tau_0_reconstruction}
\end{equation}
\begin{eqnarray}
\tau_{k}=&\tau_{k}^{\mathrm{approx}}-r_{k}^{\mathrm{approx}}+\nonumber\\&
+\int_{0}^{\infty}dk^{\prime}\left[\left\langle e_{k}^{>},h_{k^{\prime}}^{<}\right\rangle -\frac{\left\langle h_{0}^{>},e_{k^{\prime}}^{<}\right\rangle +\left\langle e_{0}^{>},h_{k^{\prime}}^{<}\right\rangle }{\left\langle h_{0}^{>},e_{0}^{<}\right\rangle +\left\langle e_{0}^{>},h_{0}^{<}\right\rangle }\left\langle e_{k}^{>},h_{0}^{<}\right\rangle \right]r_{k^{\prime}}.\label{eq_tau_k_reconstruction}
\end{eqnarray}
with $r_{0}^{\mathrm{approx}}$, $\tau_{0}^{\mathrm{approx}}$, $r_{k}^{\mathrm{approx}}$,
and $\tau_{k}^{\mathrm{approx}}$ the analytical approximate results
given, respectively, by equations \ref{eq_r_spp_1}, \ref{eq_t_spp_1},
\ref{eq_reflection_k_analytic}, and \ref{eq_transmission_k_analytic}. 
In the following, we will see how the approximate analytic result from
Sec.~\ref{subsec_scattering_analytical} can be recovered from a 
lowest order solution to the Fredholm equation.

\begin{figure}[h!]
\centering{}\includegraphics[width=0.8\textwidth]{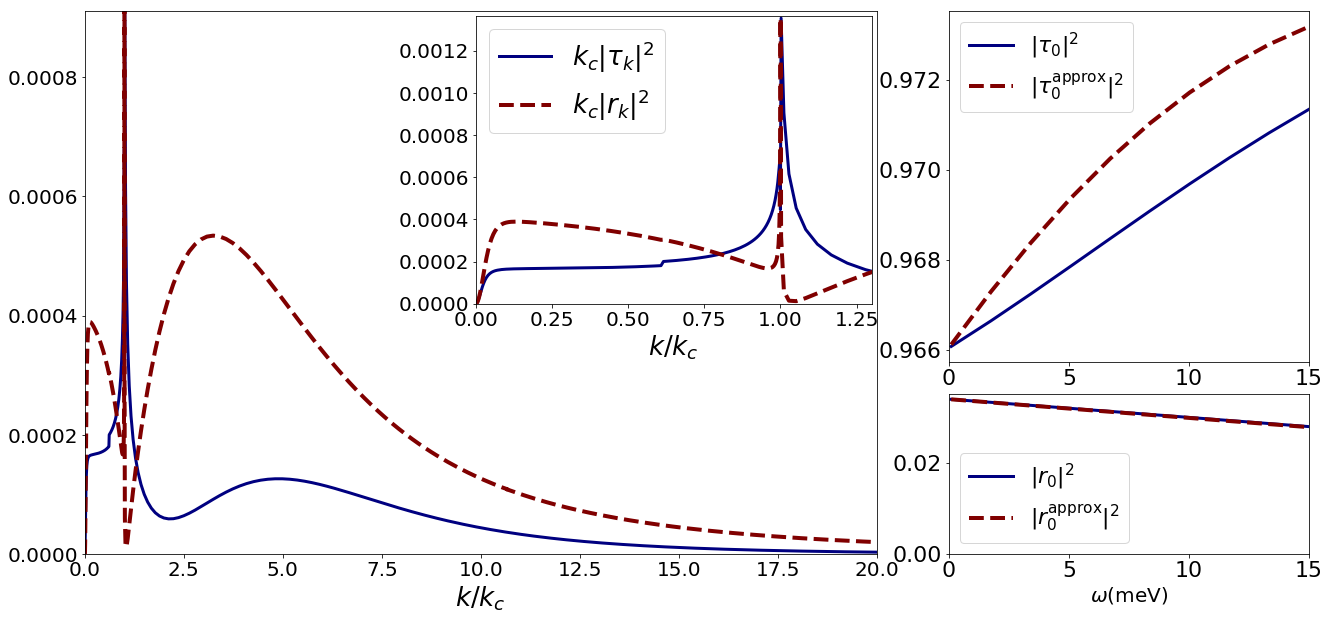}
\caption[Transmittance and reflectance of SPP]{
\label{fig_t_and_r_coefficients} 
Left panel: Transmittance and reflectance of the radiative modes as function of $k/k_c$, 
in the interval $k\in\left[0,20k_{c}\right]$, for $\omega=15$ meV. 
The inset zooms in the interval $k\in\left[0,1.25k_{c}\right]$.
Right panel: Transmittance (top right) and reflectance (bottom left) 
of graphene SPP as a function of the plasmon frequency.
Both the results obtained with the analytic approximation (dashed purple) 
and the full numerical solution (solid blue) are represented.
The difference 
between the numeric solution of Fredholm equation and the approximated solution 
for the reflection and transmission coefficients is smaller than 1\%.
In both panels the used parameters are:
meV, $d=300$ nm, $\epsilon_{1}=1.5$, $\epsilon_{2}=2.5$, $\epsilon_{3}=4$,
$E_{F}^<=0.37$ eV, $E_{F}^>=0.47$.
} \label{fig_transmittance_reflectance}
\end{figure}

\subsubsection{Recovery of the approximate analytical solution}\label{subsubsec_recovery_analytical}

We will now see how to recover the analytic result of 
equation~\ref{eq_reflection_k_analytic}
from the lowest order approximate solution of the Fredholm 
equation \ref{eq_Fredholm_second}. 
A possible strategy to solve the Fredholm equation, 
is to employ an iterative method. Within this solution scheme, the 
zeroth order solution is given by (see equation \ref{eq_Fredholm_second})
\begin{equation}
r_{k}^{(0)} = - \frac{z_{1}(k)}{v(k)}. \label{eq_zeroth_order_Fredholm}
\end{equation}
Now we notice that for $\varepsilon_{1}\simeq\varepsilon_{2}$
and $E_{F}^<\simeq E_{F}^>$, the quantity $v(k)$ can be approximated as (see \ref{app_inner})
\begin{equation}
v(k)\simeq-\frac{2}{\left\langle h_{k}^{>},e_{0}^{<}\right\rangle +\left\langle e_{k}^{>},h_{0}^{<}\right\rangle }.\label{eq_vk_approx}
\end{equation}
Therefore, we can write the reflection coefficient as
\begin{equation}
r_{k}^{(0)}\simeq\frac{1}{2}z_{1}(k)\left(\left\langle h_{k}^{>},e_{0}^{<}\right\rangle +\left\langle e_{k}^{>},h_{0}^{<}\right\rangle \right).
\end{equation}
Using equation~\ref{eq_z1} for $z_{1}(k)$, we recover equation~\ref{eq_reflection_k_analytic},
that is, the analytical solution as the zeroth order term of the Fredholm equation:
\begin{equation}
r_{k}^{(0)}\simeq\frac{\left\langle e_{0}^{>},h_{0}^{<}\right\rangle \left\langle h_{k}^{>},e_{0}^{<}\right\rangle -\left\langle h_{0}^{>},e_{0}^{<}\right\rangle \left\langle e_{k}^{>},h_{0}^{<}\right\rangle }{\left\langle h_{0}^{>},e_{0}^{<}\right\rangle +\left\langle e_{0}^{>},h_{0}^{<}\right\rangle }.
\end{equation}
We have verified numerically that the approximation given by equation~\ref{eq_vk_approx}
holds with great accuracy even if the conditions for its derivation are violated. 
This explains the good results given by the analytic approximated solution,
even for relatively large contrast between the dielectric constants and the Fermi
energies.

\begin{figure}[h!]
\centering{}
\includegraphics[width=0.7\textwidth]{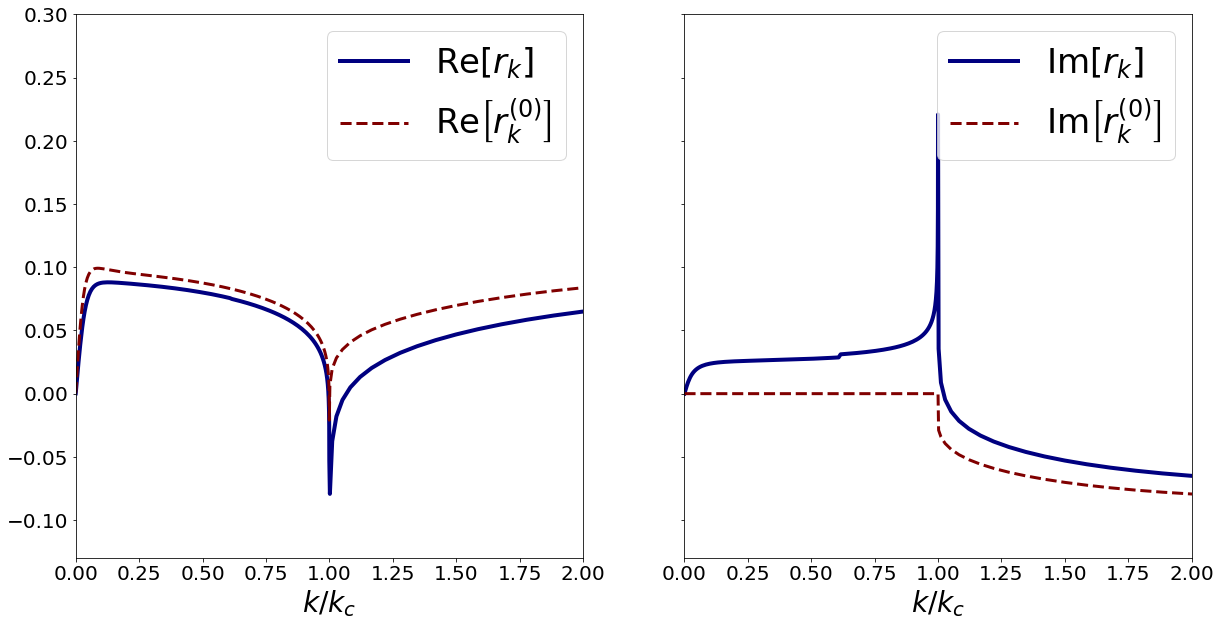} 
\caption[Solution of the Fredholm equation]{
\label{fig_Fredholm}
Real (left panel) and imaginary (right panels) parts of the reflection coefficient $r_k$ (devided by $\sqrt{d}$)
obtained from the numerical solution of the Fredholm equation (solid blue line) 
as a function of $k/k_c$ over the interval $k\in[0,2k_{c}]$. 
Also shown is the lowest order approximation to the reflection coefficient $r_k^{(0)}$ \ref{eq_zeroth_order_Fredholm} 
(dashed purple line).
The parameters used are: $\omega=15.6$ meV, $d=300$ nm,
$\epsilon_{1}=1.5$, $\epsilon_{2}=2.5$, $\epsilon_{3}=4$, $E_{F}^<=0.37$
eV, $E_{F}^>=0.47$ eV.
}
\end{figure}

\section{Results and discussion}\label{sec_results}

We shown the reflection and transmission coefficients for the SPP, $r_0$ and $\tau_0$, as a function of the plasmon
frequency, computed both with the analytic approximation (\ref{eq_r_spp_1} and \ref{eq_t_spp_1}) and with the numerical solution 
of the Fredholm equation \ref{eq_Fredholm_second} on the right panel of figure \ref{fig_transmittance_reflectance}.
As can be seen the, difference between both results is very small, not exceeding $1\%$. 
Notice however, that the approximated results overestimate the transmittance of the SPP, which is nevertheless very close to 1.
This implies that very little energy is either reflected as a SPP or lost due to emission of radiation.
This last statement is further confirmed by the smallness of the reflection and transmission coefficients 
for radiation modes as shown as a function of $k/k_c$ (with $k_c=\sqrt{\epsilon_3}k_0$)
on the left panel of figure \ref{fig_transmittance_reflectance}.  
Notice that the reflectance $\left|r_k\right|^2$ displays a significant dome for $k/k_c>1$, highlight the importance of
radiation modes evanescent along the $z$ direction in the field matching at the interface at $z=0$. In figure \ref{fig_Fredholm},
we shown the real and imaginary parts of the reflection coefficients $r_k$ obtained from the numerical solution of the Fredholm equation
and compare it to the lowest order solution as a function of $k/k_c$. The agreement is reasonable for the real part, indicating that the 
approximate analytic expressions indeed provide good results. However, in the imaginary part of the reflection coefficients
there is a significant discrepancy close to $k=k_c$, with the numerical result displaying there a peak that is absent on the approximate result.

The validity of both the analytic results and the numerical solution can be accessed by studying the total scattered, 
including the energy carried by the transmitted and reflected SPP and the energy radiated in the scattering process.
As a matter of fact, energy conservation implies that $S=1$ (see \ref{app_energy_conservation}), where
\begin{equation}
S=\vert r_{0}\vert^{2}+\vert\tau_{0}\vert^{2}+{\cal R}_{R}+{\cal T}_{R},\label{eq_energy_sum}
\end{equation}
with
\begin{eqnarray}
{\cal R}_{R} & =\int_{0}^{k_{c}}\vert r_{k}\vert^{2}dk,\label{eq_reflectance_radiation}\\
{\cal T}_{R} & =\int_{0}^{k_{c}}\vert\tau_{k}\vert^{2}dk,\label{eq_transmitance_radiation}
\end{eqnarray}
respectively, the energy radiated fraction of energy in reflection and transmission. Notice that the integration only goes up to $k_c$,
since modes with $k>k_c$ are evanescent along the $z$ direction, not carrying energy away for $z \rightarrow \pm \infty$. 
The statement $S=1$ simply means that the energy of the incident SPP is redistributed
into the reflected and transmitted SPP modes and into radiation modes

Notice that the approximate analytic results in the limits of $\omega/c \rightarrow0$ and  $d\rightarrow0$, 
\ref{eq_electro_r} and \ref{eq_electro_t}, imply that
that $\left|r_{0}\right|^{2}+\left|\tau_{0}\right|^{2}=1$. 
This means that in this is limit all the energy is carried by the transmitted and reflected SPP, with no radiation emission.
This is expected as in the electrostatic
limit no radiation can be emitted. However, in the limit of $\omega/c\rightarrow0$ and 
$d\rightarrow\infty$, equations \ref{eq_electro_r_b} and \ref{eq_electro_t_b},
imply that
\begin{equation}
\left|r_{0}\right|^{2}+\left|\tau_{0}\right|^{2}=1-\tau_{0}r_{0}^{2}.
\end{equation}
Therefore, there
is a deviation from the ideal case, $\left|r_{0}\right|^{2}+\left|\tau_{0}\right|^{2}=1$.
However, this deviation is small as long as $r_{0}\ll1$ ($\tau_{0}\lesssim1$).
We must point out, however, that the term $\tau_{0}r_{0}^{2}$ cannot
be identified with energy losses due to the emission of free
radiation, since in the electrostatic limit ($\omega/c\rightarrow0$)
the propagation of free radiation is forbidden. This deviation, is
therefore attributed to a  limitation of the approximate analytical
result. 

\begin{figure}[h!]
\centering{}
\includegraphics[width=0.7\textwidth]{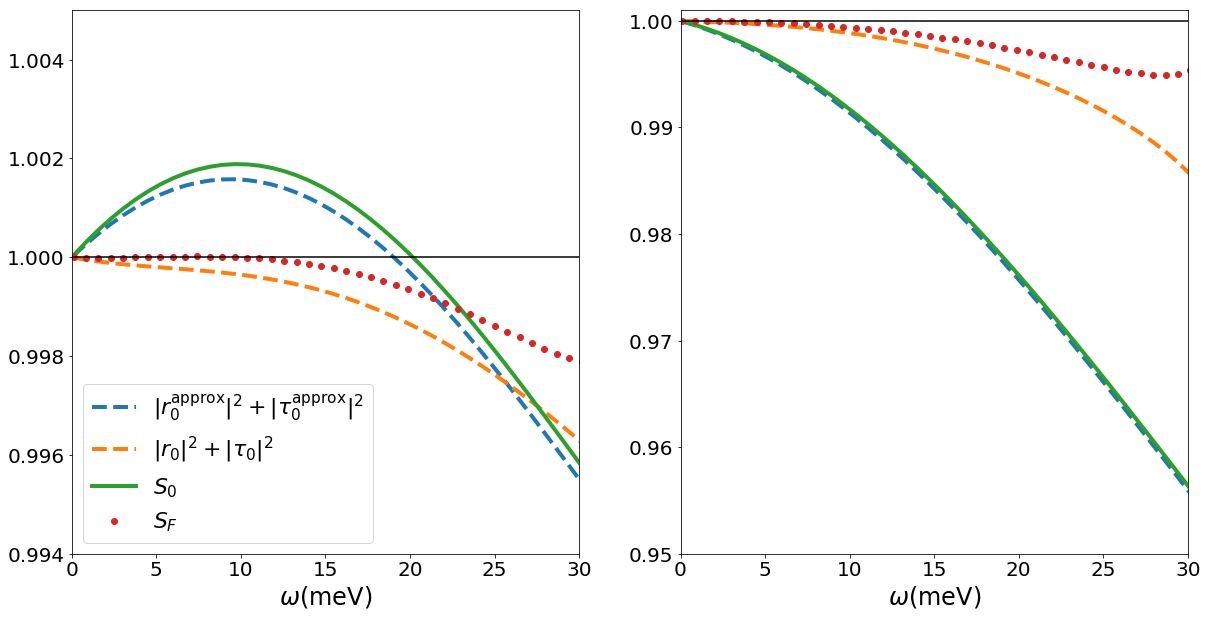}
\caption[Sum rule \ref{eq_energy_sum} in a large frequency window]{
\label{fig_sum_rule_large_freq}
Sum rule in a large frequency window.
The dashed lines refer to $\left|r_{0}\right|^{2}+\left|\tau_{0}\right|^{2}$ for the 
approximation (blue) and the numerical solution (orange). 
The green solid line refers to the approximated sum rule, $S_0$, while the 
red dotted line refers to the numerical
solution, $S_F$. The parameters used are: $d=300$ nm, $\epsilon_{1}=1.5$,
$\epsilon_{2}=2.5$, $\epsilon_{3}=4$, $E_{F}^<=0.37$ eV, $E_{F}^>=0.47$
eV. The right panel depicts the same quantities but for $E_{F}^<=0.3$
eV, $E_{F}^>=0.6$ eV. The radiative correction is the difference between the orange dashed line and the red dotted one. We can see the
increasing of radiative emission for larger frequencies and higher Fermi energy mismatch.
}
\end{figure}

To check the conservation of energy as function of frequency, of both the approximate analytic and in the numerical results, we plot
in  figure \ref{fig_sum_rule_large_freq} the
energy sum, $S$, as a function of the incident plasmon frequency in a range spanning 7.25 THz.
We see that the analytical results can violate the energy sum rule, leading to $S>1$. 
The analytical result can also lead to $\left|r^{\rm{approx}}_0\right|^2+\left|\tau^{\rm{approx}}_0\right|^2>1$, which is clearly unphysical, 
as it would correspond to a generation of energy. This indicates a limitation of the analytic approximation
which has also been reported in the scattering of surface phonon-polaritons
at the interface between two dielectrics \cite{paper_surface_phonon}. 
Notice, however, that 
the violation of the sum rule is actually very small, never exceeding $0.25\%$ (for $E_F^<=0.37$ eV and $E_F^>=0.47eV$). 
The numerical solution of the Fredholm equation corrects the unphysical result and we recover  
$\left|r^{\rm{approx}}_0\right|^2+\left|\tau^{\rm{approx}}_0\right|^2\leqslant1$. 
There is still a small violation of the sum rule which now lies bellow 1, due to errors induced by the 
discretization of the  integral in the Fredholm equation \ref{eq_Fredholm_second}. However, the numerical solution 
significantly improves the sum rule with the error being less than $0.02\%$ (for $E_F^<=0.37$ eV and $E_F^>=0.47eV$).
Notice that as we go to $\omega \rightarrow 0$ the sum rule, in both the approximate analytic 
(which completely neglects radiation modes) and in the full numerical solutions (where the contribution 
from the radiative modes is still subjected to errors due to the discretization of the integral),
is satisfied to a better degree. This is due to the fact that in the electrostatic limit the contribution due to  
radiative modes becomes less important. The errors in both methods increase when the graphene conductivity contrast 
is larger as can be seen on the  right panel of  figure \ref{fig_sum_rule_large_freq} 
(results obtained for $E_F^<=0.3$ eV and $E_F^>=0.6eV$). 
Since the sum rule is not exactly one, the fraction of energy emitted as radiation can be obtained from 
$\mathcal{R}+\mathcal{T}=S-\left|r_0\right|^2-\left|\tau_0\right|^2$, and can be seen to be extremely small,
but increases as the energy of the incident SPP increases and the graphene conductivity contrast is larger.

\section{Conclusions}\label{sec_conclusions}

We have analyzed in detail the scattering of graphene surface plasmon-polaritons
at a sharp graphene conductivity step and/or change of the dielectric
substrate. One of the merits of our calculation is the ability
to provide analytic expressions for the reflectance and transmittance
amplitudes for arbitrary values of the graphene sheet conductivity
and of the surrounding dielectric constants, in a realistic geometric
configuration. Although the analytical approach is not exact, it is
good enough to estimate the values of $r_0$ and $\tau_0$, which can
be corrected either by an iterative solution or a fully numerical solution (see \ref{numerical_solution})
of the Fredholm equation.
The corrections are, however, small. The calculation also predicts that the 
emission of free radiation in the scattering event is small. 
This situation is rather favorable for plasmon scattering,
as most of the energy remains in the plasmon field and is 
not lost to the radiation continuum. 

Note that our calculations are realistic
in what concerns the geometry of the system, since the metallic gate
is taken into account as is the existence of two different dielectrics
underneath graphene.  
However, we assumed that the induced change of the graphene conductivity is abrupt at the interface. 
A more realistic situation would be to consider a smooth transition
of the electronic density across the interface. In this case,
the reflection coefficients are no-longer well defined, except faraway
from the region where the conductivity changes; this renders the calculation
much more difficult. Nevertheless, our results should remain valid provided 
the incident plasmon wavelength  is much larger than the length scale over 
which the graphene conductivity changes. 

The method employed in this paper can be extended to
take into account the coupling of the SPP to the substrate's surface
optical phonons, as for example in SiO$_{2}$, by taking into account the 
frequency dependence of the dielectric function of the substrate. 
It is also possible to generalize the present method to a geometry
where a finite dielectric is sandwiched between two semi-infinite
ones. In this setup, by adjusting the length of the central dielectric it is possible
to achieve either total transmission or total reflection via Fabry-Pérot
oscillations, thus allowing the construction of a Bragg reflector. 
Alternatively, we can change the value of the gate potential, thus tuning the frequency
for which there is total reflection or total transmittance. This give
us a real time and on-demand control on the scattering of the plasmon. 
We point out that we have only focused on the case of scattering at normal
incidence. However, the method of eigenmode field expansion and matching
employed in this work can also be generalized and applied to the case
of oblique incidence. That extension will be the goal of a forthcoming
publication. 

\section*{Acknowlegements}

A. J. C. acknowledge the scholarship from the Brazilian agency CNPq
(Conselho Nacional de Desenvolvimento Científico e Tecnológico). 
B. A., Y. V. B. and N. M. R. P. acknowledge support from the European
Commission through the project ``Graphene-Driven Revolutions in ICT
and Beyond\char`\"{} (Ref. No. 696656). 
P.~A.~D.~Gon\c{c}alves acknowledges 
financial support from the Center for Nanostructured Graphene, 
funded by the Danish National Research Foundation (project DNRF103).
N. M. R. P. also acknowledges
the hospitality of the MackGraphe Center, at Mackenzie Presbyterian
University, where this work has started, the projects Fapesp 2012/50259-8
and 2016/11814-7, and the Portuguese Foundation for Science and Technology
(FCT) in the framework of the Strategic Financing UID/FIS/04650/2013.

\appendix

\section[Maxwell Equation]{Eigenmodes of Maxwell's equations}\label{maxwell_solution}

In this apendix we determine the eigenmodes of the system represented in figure \ref{fig_configuration_1} 
for each to the regions $z\lessgtr0$, by solving Maxwell's equations
in this geometry. The electric, $\mathbf{E}$, and the magnetic $\mathbf{B}$, fields are governed by the inhomogeneous Maxwell equations
\begin{eqnarray}
\nabla\times\mathbf{E} & =-\frac{\partial\mathbf{B}}{\partial t},\label{eq_Maxwell_curl_E}\\
\nabla\times\mathbf{B} & =\frac{\varepsilon\left(x,z\right)}{c^{2}}\frac{\partial\mathbf{E}}{\partial t}+\mu_{0}\mathbf{j},\label{eq_Maxwell_curl_B}
\end{eqnarray}
where $\mathbf{j}$ is the current density due to the graphene layer at $x=0$ and $\epsilon(x,z)$ is takes into account the inhomogeneous dieletric 
environment that surrounds the graphene layer. $\epsilon(x,z)$ is piecewise homogeneous and we write it as $\epsilon(x,z)=\epsilon^{<}(x)$ for $z<0$ 
and  $\epsilon(x,z)=\epsilon^{>}(x)$ for $z>0$, with
\begin{eqnarray}
\varepsilon^{<}\left(x\right)=\left\{ \begin{array}{@{\kern2.5pt}lL}
\hfill\epsilon_{3},& $x>0$\\
\hfill\epsilon_{1},& $-d<x<0$
\end{array}\right.,\\
\varepsilon^{>}\left(x\right)=\left\{ \begin{array}{@{\kern2.5pt}lL}
\hfill\epsilon_{3},& $ x>0$\\
\hfill\epsilon_{2},& $-d<x<0$
\end{array}\right..
\end{eqnarray}
The graphene current density is related to the eletric field by $\mathbf{j}=\sigma(z)\mathbf{E}_{\perp}$, where $\mathbf{E}_{\perp}$ represents 
the components of $\mathbf{E}$ that are perpendicular to the $x$ direction.  We also allow for different graphene conductivities
(due to different local doping levels) for $z<0$ and $z>0$, respectively, $\sigma_{<}$ and $\sigma_{>}$. 
We will use the Drude model for the graphene conductivity, namely%
\begin{equation}
\sigma{}^{\lessgtr}=\frac{e^{2}}{\pi\hbar}\frac{E_{F}^{\lessgtr}}{\gamma^{\lessgtr}+i\hbar\omega}
\end{equation}
with $E_{F}^{\lessgtr}$ the local Fermi level and $\gamma^{\lessgtr}$
the local decay rate. 

We will consider that all fields have a harmonic time dependence
of the form $e^{i\omega t}$ and also assume that the system is translationally
invariant along the $y$ direction. We want to describe scattering at normal incident and therefore we can drop all depence of the problem on
the $y$ coordinate (i.e. $\partial/\partial y=0$).
The total electromagnetic field can, in general, be split in two polarisations:
$s$/TE (transverse electric) polarization and $p$/TM (transverse
magnetic) polarization. Since the SPPs are TM-polarized waves, further
in the appendix we restrict our consideration to that particular polarization.
For this polarization and at normal incidence, the electric field will have non-zero $x-$ and
$z-$components, $\mathbf{E}=\left( E_{x},\,0,\,E_{z} \right) $,
while the only nonzero component of the magnetic field is the $y-$component, 
$\mathbf{B}=\left( 0,\,B_{y},\,0\right) $. Under these conditions
we rewrite Maxwell's equations (\ref{eq_Maxwell_curl_E}) and (\ref{eq_Maxwell_curl_B})
as 
\begin{eqnarray}
\frac{\partial E_{z}}{\partial x}-\frac{\partial E_{x}}{\partial z}=i\omega B_{y},\label{eq:Maxwell-TM-By}\\
-\frac{\partial B_{y}}{\partial z}=i\frac{\omega\varepsilon\left(x,z\right)}{c^{2}}E_{x},\label{eq:Maxwell-TM-Ex}\\
\frac{\partial B_{y}}{\partial x}=i\frac{\omega\varepsilon\left(x,z\right)}{c^{2}}E_{z}+\mu_{0}\delta\left(x\right)\sigma\left(z\right)E_{z}.\label{eq:Maxwell-TM-Ez}
\end{eqnarray}

Due to the piecewise homogenity of the system along the $z-$direction, we can study separately
the electromagnetic fields in the regions $z<0$ and $z>0$.  In general, there is a 
series of solutions, which we will refer to as eigenmodes, indexed by
some label $n$ for each of the regions $z\lessgtr0$. A general solution for each region can be
represented as a superposition of these eigenmodes. In particular,
the expression for the $y$-component of the magnetic field at $z\lessgtr0$
have the form
\begin{equation}
B_{y}^{\lessgtr}(x,z)=\sum_{n} \alpha_{n,\lambda} e^{\lambda iq_{n}^{\lessgtr}z} h_{n}^{\lessgtr}(x),\label{eq_expansion_B}
\end{equation}
while the nonzero components of the electric field are
\begin{eqnarray}
E_{x}^{\lessgtr}(x,z) & =-\sum_{n,}\pm \alpha_{n,\lambda} e^{\lambda iq_{n}^{\lessgtr}z} e_{n}^{\lessgtr}(x),\label{eq_expansion_E_x}\\
E_{z}^{\lessgtr}(x,z) & =\sum_{n} \alpha_{n,\lambda} e^{\lambda iq_{n}^{\lessgtr}z} \mathcal{E}_{n}^{\lessgtr}(x).\label{eq:expansion_E_z}
\end{eqnarray}
$\alpha_{n,\lambda}$ are the eigenmode amplitudes and the summation is taken with respect to the eigenmode index $n$. 
The $\lambda=\pm1$ sign stands for the left/right
propagating waves in $z$ direction with wavenumber $q_{n}^{\lessgtr}$.
With some abuse of notation, the summation symbol in equations~\ref{eq_expansion_B}\textendash \ref{eq:expansion_E_z}
actually represents a summation, an integral or both, depending if
the basis is discrete and/or continuous.

From equations \ref{eq:Maxwell-TM-By}-\ref{eq:Maxwell-TM-Ez}, for each mode the functions $h_n^{\lessgtr}$, $e_n^{\lessgtr}$, 
and $\mathcal{E}_n^{\lessgtr}$ are solutions of the equations
\begin{eqnarray}
\frac{\partial\mathcal{E}_{n}^{\lessgtr}}{\partial x}+iq_{n}^{\lessgtr}e_{n}^{\lessgtr}(x)=i\omega h_{n}^{\lessgtr}(x),\label{eq:Maxwell-TM-By-mode}\\
iq_{n}^{\lessgtr}h_{n}^{\lessgtr}(x)=i\frac{\omega\varepsilon^{\gtrless}\left(x\right)}{c^{2}}e_{n}^{\lessgtr}(x),\label{eq:Maxwell-Tm-Ex-mode}\\
\frac{\partial h_{n}^{\lessgtr}}{\partial x}=\left[i\frac{\omega\varepsilon^{\gtrless}\left(x\right)}{c^{2}}+\mu_{0}\delta\left(x\right)\sigma^{\gtrless}\right]\mathcal{E}_{n}^{\lessgtr}(x).\label{eq:Maxwell-Tm-Ez-mode}
\end{eqnarray}
As before, the piecewise-homogenity 
of equations \ref{eq:Maxwell-TM-By-mode}\textendash \ref{eq:Maxwell-Tm-Ez-mode} along the $x$ direction
allows us to solve them separately in regions $-d<x<0$ and $x>0$
and then apply the boundary conditions. Thus, in the region $x>0$,
occupied by the dielectric $\epsilon_{3}$ substitution of equations \ref{eq:Maxwell-Tm-Ex-mode}
and \ref{eq:Maxwell-Tm-Ez-mode} into equation~\ref{eq:Maxwell-TM-By-mode}
results into the wave equation
\begin{equation}
\frac{d^{2}h_{n}^{\lessgtr}(x)}{dx^{2}}=\left(p_{3||n}^{\lessgtr}\right)^{2}h_{n}^{\gtrless}(x),\label{eq_TM_bulk_wave_eq3}
\end{equation}
In the same way, for region $-d<x<0$, we obtain the wave equations
\begin{eqnarray}
\frac{d^{2}h_{n}^{<}(x)}{dx^{2}} & = & \left(p_{1|n}\right)^{2}h_{n}^{<}(x),\label{eq_TM_bulk_wave_eq1}\\
\frac{d^{2}h_{n}^{>}(x)}{dx^{2}} & = & \left(p_{2|n}\right)^{2}h_{n}^{>}(x),\label{eq_TM_bulk_wave_eq2}
\end{eqnarray}
which are valid for the domains $z<0$ and $z>0$,
respectively. In equations \ref{eq_TM_bulk_wave_eq3}\textendash \ref{eq_TM_bulk_wave_eq2}
$\left(p_{1|n}\right)^{2}=\left(q_{n}^{<}\right)^{2}-\epsilon_{1}k_{0}{}^{2}$,
$\left(p_{2|n}^{>}\right)^{2}=\left(q_{n}^{>}\right)^{2}-\epsilon_{2}k_{0}^{2}$,
and $\left(p_{3|n}^{\lessgtr}\right)^{2}=\left(q_{n}^{\lessgtr}\right)^{2}-\epsilon_{3}k_{0}^{2}$, 
with $k_{0}=\omega/c$ the wavenumber in vacuum. Notice that $q_{n}^{\lessgtr}$
is the same in both $x>0$ and $0>x>-d$ regions. The fact that we
have a perfect metal at $x=-d$ forces the $z-$component of the electric
field to become null there. Therefore, the magnetic field mode along
the $y$ component must have the following form
\begin{equation}
h_{n}^{\lessgtr}(x)  =
\left\{ \begin{array}{@{\kern2.5pt}lL}
    \hfill  B_{n}^{\lessgtr}e^{p_{3||n}^{\lessgtr}x}+C_{n}^{\lessgtr}e^{-p_{3||n}^{\lessgtr}x},  &  $x>0$ \\
    \hfill  A_{n}^{\lessgtr}\cosh\left[p_{j|n}\left(x+d\right)\right]   & if $0>x>-d$,
\end{array}\right. \label{eq_B_field_graphene_open}
\end{equation}
with the $x-$component of the electric field given by
\begin{equation}
e_{n}^{\lessgtr}(x)= \frac{q_{n}^{\lessgtr}c^{2}}{\omega\epsilon_{3}} \left\{ \begin{array}{@{\kern2.5pt}lL} 
\hfill B_{n}^{\lessgtr}e^{p_{3||n}^{\lessgtr}x}+C_{n}^{\lessgtr}e^{-p_{3||n}^{\lessgtr}x} & if $x>0$\\
\hfill \frac{\epsilon_{3}}{\epsilon_{j}}A_{n}^{\lessgtr}\cosh\left[p_{j|n}\left(x+d\right)\right] & if $0>x>-d$,
\end{array}\right. \label{eq_E_x_field_graphene_open}
\end{equation}
and the $z-$component being given by
\begin{equation}
\mathcal{E}_{n}^{\lessgtr}(x)=\frac{p_{3|n}^{\lessgtr}c^{2}}{i\omega\epsilon_{3}} \left\{ \begin{array}{@{\kern2.5pt}lL} 
\hfill B_{n}^{\lessgtr}e^{p_{3||n}^{\lessgtr}x}-C^{\lessgtr}e^{-p_{3||n}^{\lessgtr}x} & if $x>0$\\
\hfill \frac{p_{j|n}\epsilon_{3}}{p_{3|n}^{\lessgtr}\epsilon_{j}}A_{n}^{\lessgtr}\sinh\left[p_{j||n}\left(x+d\right)\right] & if $0>x>-d$,
\end{array}\right..\label{eq_E_z_field_graphene_open}
\end{equation}
Notice, that in equations \ref{eq_B_field_graphene_open}\textendash \ref{eq_E_z_field_graphene_open}
the subscript $j=2$ is for $z>0$ (and is combined with the superscript
$>$), while the subscript $j=1$ is for $z<0$ (combined with the
superscript $<$). Also $A_{n}^{\lessgtr}$, $B_{n}^{\lessgtr}$ and
$C_{n}^{\lessgtr}$ are coefficients to be determined such that boundary
conditions at $x=0$ are satisfied and the mode is normalized. Integration
of equations \ref{eq:Maxwell-TM-By-mode}-\ref{eq:Maxwell-Tm-Ez-mode}
in the limits from $x=0^{-}$ to $x=0^{+}$ imposes the following
boundary conditions at $x=0$
\begin{eqnarray}
\mathcal{E}_{n}^{\lessgtr}(0^{+})-\mathcal{E}_{n}^{\lessgtr}(0^{-}) & =0,\label{eq_boundary_1}\\
b_{n}^{\lessgtr}(0^{+})-b_{n}^{\lessgtr}(0^{-}) & =\mu_{0}\sigma_{\lessgtr}\mathcal{E}_{n}^{\lessgtr}(0),\label{eq_boundary_2}
\end{eqnarray}
which translate into the following equations for $A_{n}^{\lessgtr}$,
$B_{n}^{\lessgtr}$ and $C_{n}^{\lessgtr}$
\begin{eqnarray}
B_{n}^{\lessgtr}-C_{n}^{\lessgtr}  =\frac{p_{j|n}\epsilon_{3}}{p_{3|n}^{\lessgtr}\epsilon_{j}}A_{n}^{\lessgtr}\sinh\left(p_{j|n}d\right),\label{eq_graphene_open_boundary_E}\\
B_{n}^{\lessgtr}+C_{n}^{\lessgtr}-A_{n}^{\lessgtr}\cosh\left(p_{j|n}d\right)  =\frac{\sigma_{\lessgtr}}{i\omega\epsilon_{0}}\frac{p_{j|n}}{\epsilon_{j}}A^{\lessgtr}\sinh\left(p_{j|n}d\right).\label{eq_graphene_open_boundary_B}
\end{eqnarray}
By solving these equations for $B_n^{\lessgtr}$ and $C_n^{\lessgtr}$ we obtain equations \ref{eq_radiative_mode_B_coefficient} and \ref{eq_radiative_mode_C_coefficient} of the main text. 

The normalization condition \ref{eq_normalization_delta} allows to fix the value of $A_n^{\lessgtr}$. By using the following results
\begin{eqnarray}
\frac{2}{\pi}\int_{0}^{+\infty}dx\cos\left(kx\right)\cos\left(k^{\prime}x\right)  =\delta\left(k-k^{\prime}\right),\\
\frac{2}{\pi}\int_{0}^{+\infty}dx\sin\left(kx\right)\sin\left(k^{\prime}x\right)  =\delta\left(k-k^{\prime}\right),\\
\int_{0}^{+\infty}dx\cos\left(kx\right)\sin\left(k^{\prime}x\right)  =\frac{k^{\prime}}{\left(k^{\prime}\right)^{2}-k^{2}},
\end{eqnarray}
we obtain equation \ref{eq_normalization_radiative} of the main text.

\section[Energy conservation]{Energy sum rule}\label{app_energy_conservation}
Energy propagation is intimately related to the time average of the Poynting
vector $\mathbf{S}$, defined as 
\begin{equation}
\mathbf{S}=\frac{1}{2\mu_{0}}\mathbf{E}\times\mathbf{B}^{*}.
\end{equation}
For a TM-polarized electromagnetic field propagating along the $z-$direction, the Poynting vector has the explicit form
\begin{equation}
\mathbf{S}=\frac{1}{2\mu_{0}}\left(E_{x}B_{y}^{*}\mathbf{u}_{z}-E_{z}B_{y}^{*}\mathbf{u}_{x}\right).
\end{equation}
In the presence of an imaginary-only conductivity energy is conserved
and Poynting's theorem establishes that 
\begin{equation}
\mathrm{Re}\int_{\partial V}\mathbf{S}\cdot d\mathbf{A}=0,\label{eq_teorP}
\end{equation}
where $\partial V$ is the closed surface enclosing the volume $V$
and $d\mathbf{A}$ is an infinitesimal areal vector lying on the surface
of $\partial V$ and pointing from the inside to the outside of the
volume $V$. We are interested in the fields in the far-field, therefore
we draw a cube passing through $z=\pm\infty$, $x=-d$, $x=+\infty$,
and $y=\pm\infty$. As the fields do not depend on $y$ the integral
over $\partial V$ can be reduced to a one-dimensional integral along
the rectangle defined by $z=\pm\infty$, $x=-d$, and $x=+\infty$.
We now use equations \ref{eq_E_less}-\ref{eq_Bgreater} to compute
the Poynting vector. The energy flow along the $z$ direction is related
to $2\mu_{0}S_{z}^{<}(x,z)$ which reads 

\begin{eqnarray}
2\mu_{0}S_{z}^{<}(x,z)=E_{x}^{<}(x,z)B_{y}^{<*}(x,z)= \nonumber \\
=e_{0}^{<}(x)\left(e^{-iq_{0}^{<}z}-r_{0}e^{iq_{0}^{<}z}\right)h_{0}^{<*}(x)\left(e^{iq_{0}^{<}z}+r_{0}^{*}e^{-iq_{0}^{<}z}\right) \nonumber \\
-\int_{0}^{\infty}dk\int_{0}^{\infty}dk^{\prime}r_{k}r_{k^{\prime}}^{*}e_{k}^{<}(x)h_{k^{\prime}}^{<*}(x)e^{i(q_{k}^{<}-q_{k^{\prime}}^{<*})z} \nonumber \\
-\int_{0}^{\infty}dk\,r_{k}e_{k}^{<}(x)e^{iq_{k}^{<}z}h_{0}^{<*}(x)\left(e^{iq_{0}^{<}z}+r_{0}^{*}e^{-iq_{0}^{<}z}\right) \nonumber \\
+\int_{0}^{\infty}dk\,r_{k}^{*}h_{k}^{<*}(x)e^{-iq_{k}^{<*}z}e_{0}^{<}(x)\left(e^{-iq_{0}^{<}z}-r_{0}e^{iq_{0}^{<}z}\right).
\end{eqnarray}
Integrating $2\mu_{0}S_{z}^{<}(x,z)$ along the $x-$axis from $x=-d$
to $x=\infty$ and using the orthonormality of the modes it follows
that 
\begin{eqnarray}
2\mu_{0}\mathrm{Re}\int_{-d}^{\infty}dx\left[S_{z}^{<}(x,z\rightarrow-\infty)\right]=(1-\vert r_{0}\vert^{2}) \nonumber \\
-\int_{0}^{k_{c}}dk\vert r_{k}\vert^{2}-\int_{k_{c}}^{\infty}dk\vert r_{k}\vert^{2}e^{2\vert q_{R}^{<}\vert z}.
\end{eqnarray}
In the far field $z\rightarrow-\infty$ the last term of the previous
equation is zero. In the same way the contribution from the surface
located at $z=+\infty$ provides the result: 
\begin{equation}
2\mu_{0}\mathrm{Re}\int_{-d}^{\infty}dxS_{z}^{>}(x,z\rightarrow\infty)=|\tau_{0}|^{2}+\int_{0}^{k_{c}}dk|\tau_{k}|^{2}.
\end{equation}
Finally, we still need the contribution from the line at $x=+\infty$.
The last term we need to compute is: 
\begin{equation}
\mathrm{Re}\int_{-\infty}^{\infty}dzS_{x}(x,z)=\mathrm{Re}\int_{-\infty}^{0}dzS_{x}^{<}(x,z)
+\mathrm{Re}\int_{0}^{\infty}dzS_{x}^{>}(x,z),
\end{equation}
which corresponds to radiation emitted orthogonal to the graphene
plane. The plasmonic fields $e_{0}$ and $h_{0}$ go to zero when
$x\rightarrow\infty$ and thus do not transport energy. Therefore we are
left with the term that depends on the radiative modes. It can be
shown that the integral is purely imaginary and therefore its real
part is zero and does not contributes to energy conservation. Putting
all together in equation (\ref{eq_teorP}) we find 
\begin{equation}
1=\left|\tau_{0}\right|^{2}+\left|r_{0}\right|^{2}+\int_{0}^{k_{c}}dk\left|r_{k}\right|^{2}+\int_{0}^{k_{c}}dk\left|\tau_{k}\right|^{2},
\end{equation}
which is the statement of energy conservation.

\section[Explicit form of the inner products]{Explicit form of the inner products}\label{app_inner}

In this Appendix we list the explicit results for the inner products.
First we provide results for some useful integrals: 
\begin{eqnarray}
\int_{-d}^{0}dx\cosh(p_{j}(x+d))\cosh(p_{j^{\prime}}(x+d))=\nonumber\\
=\frac{1}{2}\left[\frac{\sinh(p_{j}+p_{j^{\prime}})d}{p_{j}+p_{j}^{\prime}}+\frac{\sinh\left[(p_{j}-p_{j}^{\prime})d\right]}{p_{j}-p_{j^{\prime}}}\right],
\end{eqnarray}
\begin{eqnarray}
\int_{0}^{\infty}dxe^{-2p_{3}x} & =\frac{1}{2p_{3}},\\
\int_{0}^{\infty}dxe^{-p_{3}x}\sin(kx) & =\frac{k}{k^{2}+p_{3}^{2}},\\
\int_{0}^{\infty}dxe^{-p_{3}x}\cos(kx) & =\frac{p_{3}}{k^{2}+p_{3}^{2}},
\end{eqnarray}
and 
\begin{equation}
\int_{0}^{\infty}dx\cos(kx)\sin(k^{\prime}x)e^{-0^{+}x}=\frac{1}{2}\left(\frac{1}{k+k^{\prime}}+\frac{1}{k^{\prime}-k}\right)\,.
\end{equation}
Using the previous integrals we can compute the different inner products,
which, after tedious calculations, read:
\be
\langle e_0^<,h_0^>\rangle=  q^< \tilde{A}_1 \tilde{A}_2 \left[\frac{1}{\varepsilon_1}S(p_1,p_2) +\frac{\varepsilon_3}{\tilde{p}_3^<+\tilde{p}_3^>}a_1 a_2\right],
\ee
\be
\langle h_0^<,e_0^> \rangle = q^> \tilde{A}_1 \tilde{A}_2 \left[\frac{1}{\varepsilon_2}S(p_1,p_2) +\frac{\varepsilon_3}{\tilde{p}_3^<+\tilde{p}_3^>}a_1 a_2\right],
\ee
\be
\langle e^<(k),h_0^>\rangle=  \sqrt{d}\,\tilde{q}^R \tilde{A}_1^R(\tilde{k}) \tilde{A}_2  \left( \frac{1}{\varepsilon_1}S( \tilde{p}_1^R,\tilde{p}_2)+\frac{a_1}{\tilde{k}^2+\tilde{p}_3^{>2}}T_1(\tilde{k}) \right),
\ee
\be
\langle h^<(k),e_0^>\rangle= \sqrt{d}\,\tilde{q}^> \tilde{A}_1^R(\tilde{k}) \tilde{A}_2  \left( \frac{1}{\varepsilon_2}S( \tilde{p}_1^R,\tilde{p}_2)+\frac{a_1}{\tilde{k}^2+\tilde{p}_3^{>2}}T_1(\tilde{k}) \right),
\ee
\be
\langle e_0^<,h^>(k)\rangle= \sqrt{d}\,\tilde{q}^< \tilde{A}_1 \tilde{A}_2^R(\tilde{k}) \left(\frac{1}{\varepsilon_1}S(\tilde{p}_1,\tilde{p}_2^R)+\frac{a_2}{\tilde{k}^2+\tilde{p}_3^{<2}}T_1(\tilde{k}) \right),
\ee
\be
\langle h_0^<,e^>(k)\rangle= \sqrt{d}\,\tilde{q}^R \tilde{A}_1 \tilde{A}_2^R(\tilde{k}) \left( \frac{1}{\varepsilon_2}S(\tilde{p}_1,\tilde{p}_2^R)+\frac{a_2}{\tilde{k}^2+\tilde{p}_3^{<2}}T_1(\tilde{k}) \right),
\ee
\bea
\langle e^<(k^\prime), h^>(k) \rangle= \sqrt{d} \,\tilde{q}^R(k^\prime) A_1^R(k^\prime)  A_2^R(k) \left[ \frac{1}{\varepsilon_1}
S({\tilde{p}_1}^{R\prime},\tilde{p}_2^R)+\right.\nonumber\\ \left.+ \frac{2}{\pi\varepsilon_3}Q_1(k) \delta(k-k^\prime)+ \frac{Q_2(k,k^\prime)}{k^2-{k^\prime}^2} 
\right], \label{eq_inner_delta_1}
\eea
\bea
\langle h^<(k^\prime), e^>(k) \rangle=  \sqrt{d} \, \tilde{q}^R(k) A_1^R(k^\prime)  A_2^R(k)  \left[ \frac{1}{\varepsilon_2}
S({\tilde{p}_1}^{R\prime},\tilde{p}_2^R)+\right.\nonumber\\ \left.+ \frac{2}{\pi\varepsilon_3}Q_1(k) 
\delta(\tilde{k}-\tilde{k}^\prime)+ \frac{Q_2(k,k^\prime)}{\tilde{k}^2-{(\tilde{k}^\prime)}^2} 
\right],   \label{eq_inner_delta_2}
\eea
where we have defined the functions
\be
S(x,y) = \frac{1}{2} \left( \frac{\sinh(x+y)}{x+y}+\frac{\sinh(x-y)}{x-y} \right),
\ee
\be
a_i= \frac{\cosh(\tilde{p}_3^i)}{\varepsilon_3} \left( 1-\frac{\tilde{p}_3^i \sigma_i^I}{\omega d\varepsilon_0\varepsilon_3}\right)^{-1}
\ee
where $\tilde{p}_3^1= p _3^< d$, $\tilde{p}_3^2= p_3^>d$, $\tilde{k}=kd$, and:
\be
T_i(k)= R_i(k)\tilde{p}_3^i+Z_i(k),
\ee
\be
R_i(k)=\cosh(\tilde{p}^R_i)-\frac{\sigma_i^I \tilde{p}^R_i}{\omega d \varepsilon_i \varepsilon_0} \sinh(\tilde{p}^R_i),
\ee
\be
Z_i(k)=\frac{\tilde{p}^R_i}{\varepsilon_i} \sinh(\tilde{p}^R_i),
\ee
\be
Q_1(k)=R_1(\tilde{k}) R_2(\tilde{k})+\frac{1}{\tilde{k}^2}Z_1(\tilde{k})Z_2(\tilde{k}),
\ee
\be
Q_2(k,k^\prime)=-R_1(k^\prime)\tilde{p}_2^R(k)+R_2(k) \tilde{p}_1^{R}(k^\prime),
\ee
\be
\tilde{q}^R(k)=d\sqrt{ \epsilon_3 k_0^2-k^2} ,
\ee
\be
\tilde{p}^R_i=\tilde{p}^R_i(k)=d\sqrt{\varepsilon_i k_0^2-k^2},
\ee
\be
\tilde{A}^R_i(k) =\sqrt{\frac{c^2}{\omega}} A^\lessgtr_k,
\ee
\be
\tilde{A}_i=A_0^\lessgtr,
\ee
where in the last two equations, $i=1$ ($i=2$) for the superscript $<$ ($>$).

From equations \ref{eq_inner_delta_1} and \ref{eq_inner_delta_2} and
from equation~\ref{eq_z1} and function $v(k)$, as defined by equation~\ref{eq_z3},
is given by 
\begin{equation}
v(k)=-\frac{4}{\pi\epsilon_{3}}\frac{c^{2}}{\omega}\frac{Q_{1}(k)\tilde{q}^R(k)A_{k}^{>} A_{k}^{<}  }{\left\langle h_{k}^{>},e_{0}^{<}\right\rangle +\left\langle e_{k}^{>},h_{0}^{<}\right\rangle }.\label{eq_v_function_compact}
\end{equation}
We point out that for $\varepsilon_{1}\simeq\varepsilon_{2}$ and
$E_{F}^<\simeq E_{F}^>$ we have that
\begin{equation}
\frac{4}{\pi\epsilon_{3}}\frac{c^{2}}{\omega}Q_{1}(k)\tilde{q}^R(k)A_{k}^{>} A_{k}^{<} \simeq2,
\end{equation}
such that we obtain
\begin{equation}
v(k)\simeq-\frac{2}{\left\langle h_{k}^{>},e_{0}^{<}\right\rangle +\left\langle e_{k}^{>},h_{0}^{<}\right\rangle }.
\end{equation}

\section{Numerical solution of Fredholm Problem \label{numerical_solution}}

To solve the Fredholm equation \ref{eq_Fredholm_second}, first we introduce a cutoff $k_{max}=c_F k_c$ in the integral, where $c_F$ is large 
and is choosen as the value needed for the solution to converge.
 The kernel of the Fredholm equation, $z_3(k,k')$, has a divergence of the kind:
\begin{equation}
\frac{1}{k-k'},
\end{equation}
that comes from the term proportional to $Q_2(k,k')$ in the inner products \ref{eq_inner_delta_1} and \ref{eq_inner_delta_2}. 
To regularize this divergence, we make the substitution:
\begin{equation}
\frac{1}{k-k^\prime}\rightarrow \frac{k-k'}{(k-k^\prime)^2+\eta^2},
\end{equation}
where $\eta$ is a parameter choosen as small as necessary to achieve convergence of the calculation. 
In the numerical results shown in the main text, we used $c_F=30$ and $\eta=10^{-3}k_c$.

In the integral of equation \ref{eq_Fredholm_second} we make the variable change $u=k_c k$, and separate the integration limit in two parts:
\begin{equation}
\int_0^{c_F} d u= \int_0^1 du + \int_1^{c_F}du.
\end{equation}
Next, we divide each of those integrals in $N_1$ and $N_2$ equally spaced regions. For each of those regions, we apply a Gauss-Legendre quadrature
with $N_\mathrm{Gauss1}$ (when $u<1$) and  $N_\mathrm{Gauss2}$ ($u>1$) points. The Fredholm problem now is transformed into a matrix equation:
\begin{equation}
\boldsymbol{r}=\boldsymbol{r}_0-\boldsymbol{Z}_3 \cdot \boldsymbol{r}, \label{matrix_fredholm}
\end{equation}
where $\boldsymbol{Z}_3$ is a $(N_1 N_\mathrm{Gauss1}+N_2 N_\mathrm{Gauss2})\times(N_1 N_\mathrm{Gauss1}+N_2 N_\mathrm{Gauss2})$ 
matrix obtained from the discretization of the kernel $z_3(k,k^\prime)$, $\boldsymbol{r}$ is the solution we seek, being a vector obtained by descritizing the 
reflection coefficient, and $\boldsymbol{r}_0$ is vector obtained from the discritization of the zeroth order solution of the Fredholm equation 
\ref{eq_zeroth_order_Fredholm}.
The solution of \ref{matrix_fredholm} is obtained trivially as $\boldsymbol{r}=(1+\boldsymbol{Z}_3)^{-1} \cdot \boldsymbol{r}_0$. 
For the results shown in this paper we used $N_1=N_2=80$, $N_\mathrm{Gauss1}=2$, $N_\mathrm{Gauss2}=3$.
 
This numerical procedure works for the spectral range shown in this paper 
(frequencies up to $7.25$ THz). For higher frequencies, the integration
of the resulting $\tau_k$ function, to calculate the sum rule \ref{eq_energy_sum},
diverges due to the singularity at the $k_c$ point (see figure \ref{fig_transmittance_reflectance}).
To go to higher frequencies, a more sophisticated integration algorithm is necessary.

\section*{References}
\bibliographystyle{iopart-num}
 
 
 \providecommand{\newblock}{}

\end{document}